\documentclass[aps,amssymb,amsmath,pra,twocolumn,showpacs,superscriptaddress]{revtex4-1}

\usepackage{graphicx}
\usepackage{dcolumn}
\usepackage{bm}
\usepackage{subfigure}
\usepackage{color}
\usepackage[usenames,dvipsnames]{xcolor}

\usepackage{amssymb,amsmath,amsfonts,latexsym,graphicx,verbatim,dsfont}
\usepackage[matrix,frame,arrow]{xy}
\usepackage{epsf}

\usepackage[margin=0.6in]{geometry}

\usepackage[english]{babel}
\usepackage{times}
\usepackage{latexsym}
\usepackage{fancyhdr}
\usepackage{float}
\usepackage{afterpage}
\usepackage{enumitem}
\usepackage{eso-pic, graphicx}

\definecolor{Ablue}{rgb}{0.96,0.24,0.00}

\definecolor{Abluetitle}{rgb}{0.,0.24,0.51}
\newcommand{\bluetitle}{\color{Abluetitle}}

\definecolor{orange}{rgb}{0.96,0.24,0.00}

\definecolor{darkred}{rgb}{0.55, 0.0, 0.0}

\usepackage[colorlinks=true , citecolor=blue,urlcolor=blue]{hyperref}

\newcommand{\angstrom}{\mbox{\normalfont\AA}}

\newcommand{\xg}{\gamma}
\newcommand{\xt}{\theta}

\newcommand{\xo}{\omega}

\newcommand{\app}{\approx}

\newcommand{\Cs}{{}^{13}\R{C}}

\newcommand{\xD}{\Delta}

\newcommand{\fr}[2]{\frac{#1}{#2}}

\newcommand{\sq}[1]{\sqrt{#1}}

\newcommand{\mH}[0]{\mathcal{H}}

\newcommand{\beq}{\begin{equation}}
\newcommand{\eeq}{\end{equation}}
                  
\newcommand{\benum}{\begin{enumerate}}
\newcommand{\eenum}{\end{enumerate}}
                    
\newcommand{\bit}{\begin{itemize}}
\newcommand{\eit}{\end{itemize}}

\newcommand{\bea}{\begin{eqnarray}}
\newcommand{\eea}{\end{eqnarray}}

\newcommand{\bev}{\begin{verbatim}}
\newcommand{\eev}{\end{verbatim}}

\newcommand{\zt}{\times}

\newcommand{\T}[1]{\textbf{#1}}
\newcommand{\I}[1]{\textit{#1}}
\newcommand{\R}[1]{\textrm{#1}}

\newcommand{\zl}[1]{\label{eqn:#1}}
\newcommand{\zr}[1]{Eq. (\ref{eqn:#1})}
\newcommand{\zfl}[1]{\protect\label{fig:#1}}
\newcommand{\zfr}[1]{Fig. \ref{fig:#1}}

\newcommand{\zsl}[1]{\label{sec:#1}}
\newcommand{\zsr}[1]{Sec. \ref{sec:#1}}

\newcommand{\ba}{\left\{ \begin{array}{lr}}
\newcommand{\ea}{\end{array}\right.}

\newcommand{\blist}[1]{
 \begin{list}{#1}
 \begin{align}
	 arrow
 \end{align}
 $\checkmark\star
  { \setlength{\itemsep}{3pt}
     \setlength{\parsep}{2pt}
     \setlength{\topsep}{3pt}
     \setlength{\partopsep}{0pt}
     \setlength{\leftmargin}{1em}
     \setlength{\labelwidth}{1em}
     \setlength{\labelsep}{0.5em} } }
\newcommand{\elist}{
  \end{list}  }

\DeclareMathSymbol{\vartheta}{\mathalpha}{letters}{"12}
\DeclareMathSymbol{\theta}{\mathalpha}{letters}{"23}
\DeclareMathSymbol{\phi}{\mathalpha}{letters}{"27}
\DeclareMathSymbol{\varphi}{\mathalpha}{letters}{"1E}

\newcommand{\bef}
{
\begin{figure}[htbp]
\centering
}

\newcommand{\eef}{\end{figure}}

\newcommand{\beginsupplement}{%
        \setcounter{table}{0}
        \renewcommand{\thetable}{S\arabic{table}}%
        \setcounter{figure}{0}
        \renewcommand{\thefigure}{S\arabic{figure}}%
     }

\newcommand{\affA}{Department of Chemistry, University of California Berkeley, and Materials Science Division Lawrence Berkeley National Laboratory, Berkeley, California 94720, USA.}
\newcommand{\affB}{Department of Physics, CUNY-City College of New York, New York, NY 10031, USA}
\newcommand{\affC}{CUNY-Graduate Center, New York, NY 10016, USA}
\newcommand{\affD}{Department of Chemical and Biomolecular Engineering, and Materials Science Division Lawrence Berkeley National Laboratory University of California, Berkeley, California 94720, USA.}
\newcommand{\affE}{Fakult\"{a}t Physik, Technische Universit\"{a}t Dortmund, D-44221 Dortmund, Germany}
\newcommand{\affF}{Department of Physics, University of California Berkeley, Berkeley, California 94720, USA.}
\newcommand{\affG}{Department of Physics, Peking University, Beijing, China.}

\begin{document}
\title{\bluetitle{Orientation independent room-temperature optical $\Cs$ hyperpolarization in powdered diamond}}

\author{A. Ajoy}\email{ashokaj@berkeley.edu}\affiliation{\affA}
\author{K. Liu}\affiliation{\affA}
\author{R. Nazaryan}\affiliation{\affA}
\author{X. Lv}\affiliation{\affA}
\author{P.R. Zangara}\affiliation{\affB}
\author{B. Safvati}\affiliation{\affF}
\author{G. Wang}\affiliation{\affA}\affiliation{\affG}
\author{D. Arnold}\affiliation{\affA}
\author{G. Li}\affiliation{\affA}
\author{A. Lin}\affiliation{\affA}
\author{P. Raghavan}\affiliation{\affA}
\author{E. Druga}\affiliation{\affA}
\author{S. Dhomkar}\affiliation{\affB}
\author{D. Pagliero}\affiliation{\affB}
\author{J. A. Reimer}\affiliation{\affD}
\author{D. Suter}\affiliation{\affE}
\author{C. A. Meriles}\affiliation{\affB}\affiliation{\affC}
\author{A. Pines}\affiliation{\affA}

\begin{abstract}
  
Dynamic nuclear polarization via contact with electronic spins has emerged as an attractive route to enhance the sensitivity of nuclear magnetic resonance (NMR) beyond the traditional limits imposed by magnetic field strength and temperature. Among the various alternative implementations, the use of nitrogen vacancy (NV) centers in diamond -- a paramagnetic point defect whose spin can be optically polarized at room temperature -- has attracted widespread attention, but applications have been hampered by the need to align the NV axis with the external magnetic field. Here we overcome this hurdle through the combined use of continuous optical illumination and a microwave sweep over a broad frequency range. As a proof of principle, we demonstrate our approach using powdered diamond where we attain bulk $\Cs$ spin polarization in excess of 0.25 percent under ambient conditions. Remarkably, our technique acts efficiently on diamond crystals of all orientations, and polarizes nuclear spins with a sign that depends exclusively on the direction of the microwave sweep. Our work paves the way towards the use of hyperpolarized diamond particles as imaging contrast agents for biosensing and, ultimately, for the hyperpolarization of nuclear spins in arbitrary liquids brought in contact with their surface. 
\end{abstract}

\maketitle

\zsl{DNP}

\section*{Introduction}

Nuclear Magnetic Resonance (NMR) is a widely used spectroscopic technique~\cite{Ernst}, and a true workhorse in a variety of fields ranging from chemical structure analysis to medical imaging~\cite{Wuthrich03}. In spite of its versatility and broad applicability, its inherent low sensitivity has prevented some applications of the technology, for instance in desktop spectrometers and in point-of-care medical use. Dynamic nuclear polarization (DNP) -- the ability to employ electron spins to enhance the polarization of, and hence signal from, nuclear spins~\cite{Abragam78} -- has emerged as an attractive solution for several applications. However the need to perform the electronic polarization at cryogenic temperatures and high magnetic fields~\cite{ArdenkjaerLarsen03,Maly08} has motivated the search for simpler, low-cost hyperpolarization alternatives.

A particularly compelling idea, which has garnered much recent attention, is the use of atom-like defects in diamond as \I{optical} hyperpolarizing agents~\cite{Abrams14}. Specifically, the electronic spin corresponding to the diamond Nitrogen Vacancy (NV) center is optically polarizable to $\app$ 99\% at room temperature~\cite{Jelezko06}, possesses remarkable coherence properties~\cite{Balasubramanian09}, and can be created close ($<$4nm) to the surface so as to be hyperfine coupled to external nuclei~\cite{Staudacher13,Lovchinsky16}. These attributes facilitate coherent transfer of polarization from the NV centers to proximate nuclei, boosting their NMR signal by orders of magnitude at room temperature. Indeed, large ($>$0.5\%) optical hyperpolarization of $\Cs$ nuclear spins in \I{single crystal} diamond was demonstrated recently by a variety of DNP techniques~\cite{Fischer13,Alvarez15,King15,Pagliero2018}.

\begin{figure}[t]
		 \centering
  {\includegraphics[width=0.49\textwidth]{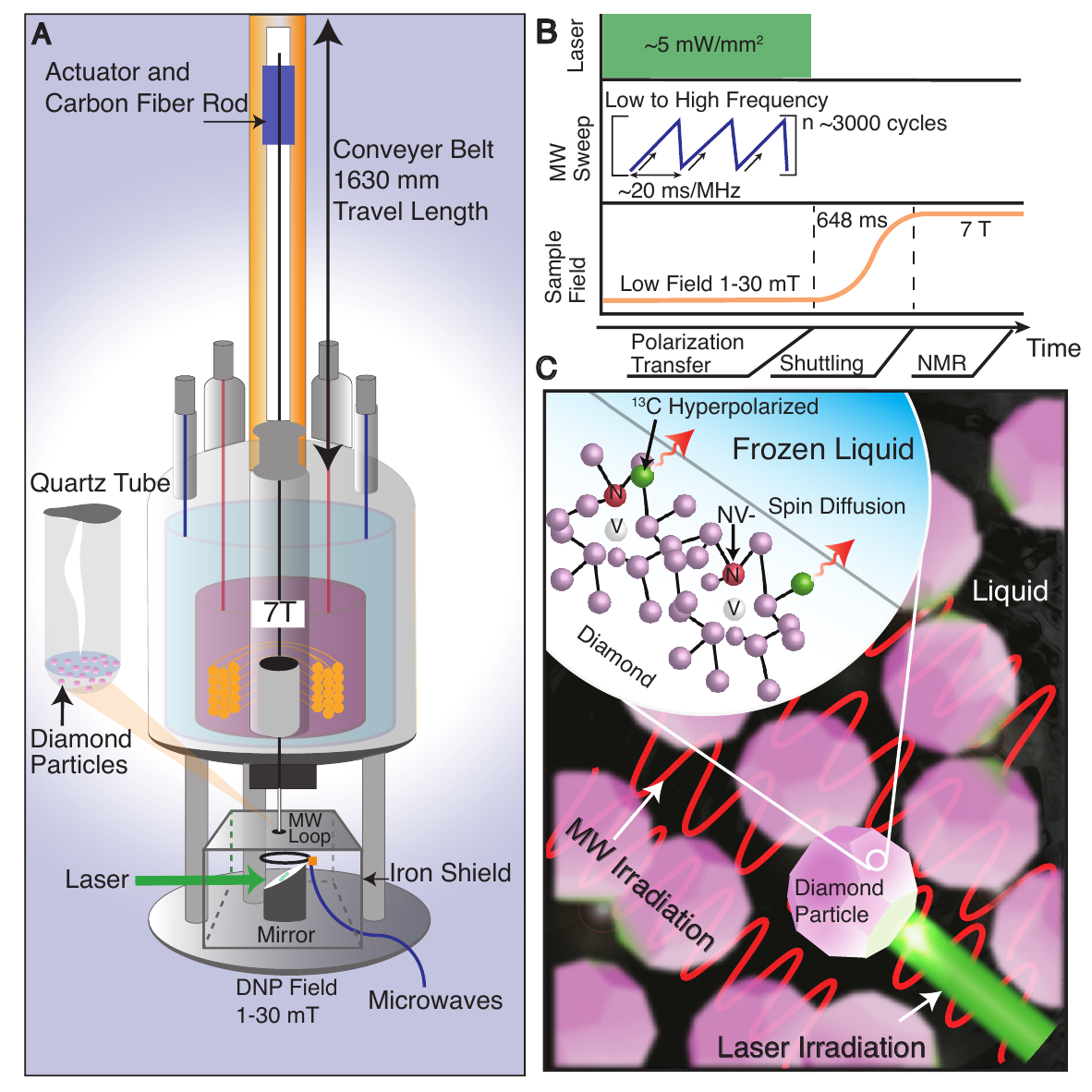}}
  \caption{\T{Experiment overview.} (A) Polarization transfer from optically pumped NV centers to $\Cs$ in diamond particles is carried out by microwave irradiation at low field ($\T{B}_{\R{pol}}\sim$ 1-30mT) after which the sample is shuttled rapidly for bulk inductive readout at 7T. We quantify the polarization enhancement with respect to the thermal signal at 7T. (B) \I{Polarization transfer protocol.} Laser light (532nm) is continuously applied along with swept microwave (MW) irradiation across the NV center spectrum at $\T{B}_{\R{pol}}$ to hyperpolarize the $\Cs$ nuclei. The sweep time per unit bandwidth is 20 ms/MHz. (C) \I{Envisioned nanodiamond polarizer.} Optically hyperpolarized $\Cs$ diamond nuclei relay polarization to $\Cs$ spins in a frozen liquid by spin diffusion aided by the intrinsically large surface area of nanoparticles. Subsequent rapid thaw would allow enhanced NMR detection with chemical shift resolution.}
\zfl{setup-main}	
\end{figure}

Despite this encouraging progress, these methods have been limited to single crystals. DNP transfer to outside spins has remained unsuccessful due to their reduced contact surface area to the external liquid. A more viable alternative is the use of diamond in powdered form, either as nano- or micro-scale particles, which offers a larger contact surface area, for instance $\gtrsim 6700 \R{mm}^2$/mg for 100nm particles, orders of magnitude greater than $\sim 0.13\R{mm}^2$/mg for a single crystal of equivalent mass. Indeed the goal of optically ``\I{hyperpolarized nanodiamonds}'' has been a long-standing one~\cite{chen15b}; yet the strong orientational dependence of the spin-1 NV centers has remained challenging to surmount~\cite{Scott16}. Unlike a single crystal with a narrow resonance, the electronic linewidth of a micro- or nanodiamonds is greatly (inhomogenously) broadened to a $\gtrsim$ 5.7GHz powder pattern even at modest fields $>$0.1T making conventional DNP strategies ineffective.

In this paper, we overcome these challenges to optically hyperpolarize diamond powder, obtaining high bulk $\Cs$ polarization comparable to the best results in single crystals~\cite{Fischer13}. Note that this is in comparison to past results wherein hyperpolarized signals were measured against Boltzmann signals under the same experimental conditions. We have developed a new, remarkably simple, \I{low-field} optical DNP technique that proves to be fully orientation independent.  Unlike conventional DNP~\cite{Maly08}, the regime in which we perform the transfer exploits the fact the NV electrons can be polarized independent of field, and low-field can be used advantageously to reduce the broadening of the electronic linewidth.  Here the bulk nuclear polarization is unambiguously detected by inductive readout subsequent to rapidly shuttling the hyperpolarized powder to high field (\zfr{setup-main}A), and is compared against the corresponding Boltzmann polarization at 7T, similar to \cite{Fischer13,Alvarez15}, but in contrast to previous work on single NV centers or small ensembles~\cite{London13,Scheuer16,Broadway17}. The DNP protocol is detailed in \zfr{setup-main}B -- nuclear hyperpolarization is affected by sweeping microwave irradiation across the NV center powder pattern at a low field $\T{B}_{\R{pol}}\app$ 1-30mT under continuous laser irradiation.

\section*{Results}

\begin{figure}[t]
  \centering
  {\includegraphics[width=0.49\textwidth]{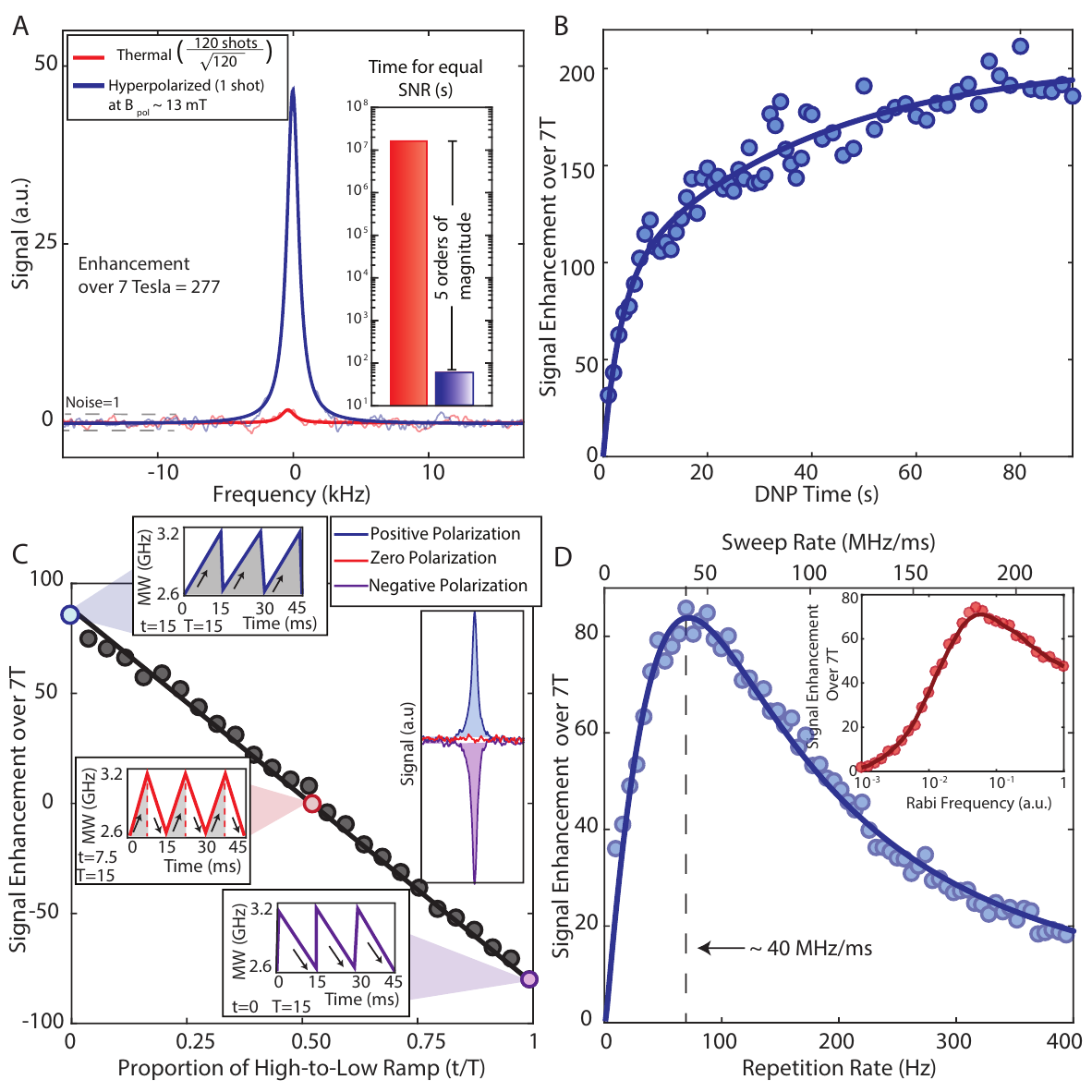}}
  \caption{\T{Optical hyperpolarization in diamond microparticles.} Hyperpolarization experiments were performed on dry 200$\mu$m particles with 1.1 \% natural abundance $\Cs$. Solid fit lines are depicted over data points. (A) \I{Signal gain by DNP} under optimized conditions. Blue line shows the $\Cs$ NMR signal due to Boltzmann polarization at 7T, averaged 120 times over 7 hours. Red line is a single shot DNP signal obtained with 60s of optical pumping, enhanced by 277 over the 7T thermal signal (enhanced 149153 times at $B_{\R{pol}}$=13mT). The signals have their noise unit-normalized for clarity. Hyperpolarization thus leads to over 5 orders of magnitude gains in averaging time (inset). (B) \I{Buildup curve} showing rapid growth of bulk $\Cs$ polarization. Slow rise at longer times is due to $\Cs$ spin diffusion. (C) \I{Hyperpolarization sign} is controlled by MW sweep direction across the NV center powder pattern. Continuous family of sweeps demonstrating the idea, with extremal points representing low-to-high frequency MW sweeps and vice-versa. Time $t$ is period of the high-to-low frequency component in one cycle of total period $T$.  \I{Inset:} $\Cs$ signal undergoes near-perfect sign inversion upon reversal of the sweep direction. Sweeping in a symmetric fashion leads to net cancellation, and no buildup of hyperpolarization. (D) \I{Sweep rate dependence} of the signal enhancement. The sweep bandwidth is 570MHz, and the excitation laser power is $\app$5mW/mm$^2$. The solid line is the result of a fit using the expression in the main text; we find $k$ = 18.4 MHz/ms and $\Lambda$ = 30 kHz for a Rabi field $\Omega$ = 0.35 MHz. \I{Inset:} Dependence of $\Cs$ NMR signal as a function of the MW Rabi frequency. Here the solid line is a guide to the eye.}	
 \zfl{results}	
\end{figure}

\begin{figure}[t]
  \centering
  {\includegraphics[width=0.47\textwidth]{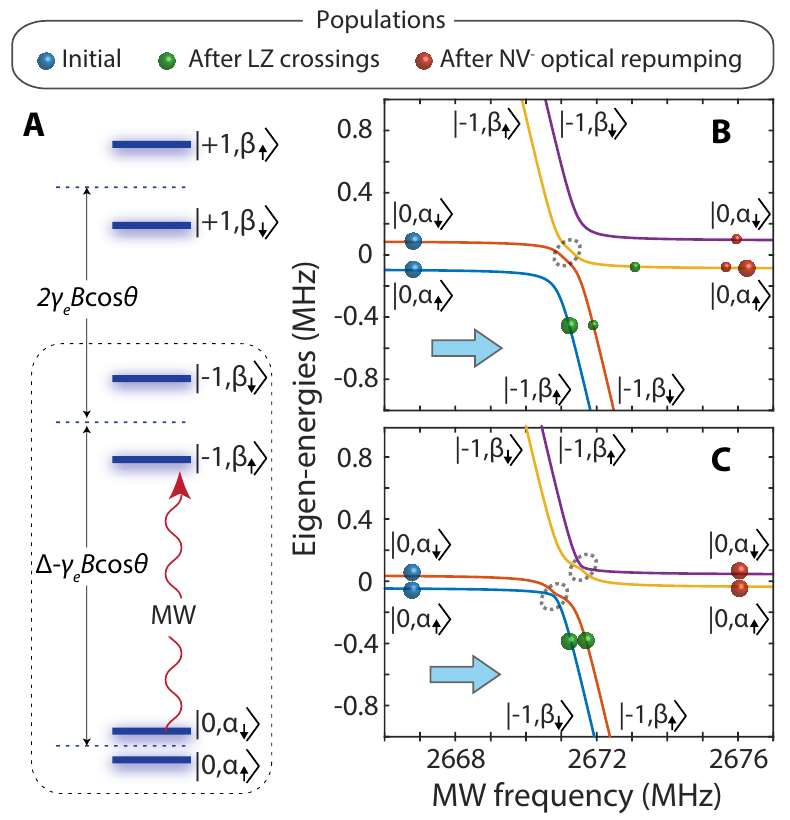}}
  \caption{
   \textbf{Proposed mechanism of polarization transfer.} (A) Energy levels of an NV\textsuperscript{-} electron spin hyperfine-coupled to a $\Cs$ nuclear spin. $\Delta$ denotes the NV zero-field splitting, $\gamma_e$ is the electron gyromagnetic ratio, $B$ (assumed much smaller than $\Delta$) is the external magnetic field forming an angle $\theta$ with the NV axis. The quantum numbers in all kets refer to electron and nuclear spins, in that order; the notation for the nuclear spin states highlights the manifold-dependent quantization axis, in general different from the magnetic field direction. (B) Calculated energy diagram in the rotating frame corresponding to the $m_S=0\leftrightarrow m_S=-1$ subset of transitions (dashed rectangle in (A)) assuming a hyperfine coupling $A_{zz}=+0.5$ MHz. (C) Same as in (B) but for $A_{zz}=-0.5$ MHz. In (B) and (C) we assume $B$=10 mT, $\theta$=45 deg., and use a transverse hyperfine constant $A_{zx}=0.3|A_{zz}|$. Colored solid circles denote populations at different stages during a sweep in the direction of the arrow, and faint dashed circles indicate the narrower avoided crossings where population transfer takes place. LZ: Landau-Zener; MW: microwave.  
	}
\zfl{mechanism}
\end{figure}


\zfr{results} summarizes the key features of the technique, demonstrated for a typical example of 200$\mu$m microparticles with 1.1 \% natural abundance $\Cs$ containing about 1ppm of NV centers (see Materials and Methods). Under optimized conditions, we obtain $\Cs$ hyperpolarization over 277 times that of the 7T Boltzmann level (\zfr{results}A) -- a high polarization level comparable to the best results on single crystals~\cite{Fischer13,Alvarez15}, yet achieved here on a completely randomly oriented powder. The use of optical pumping enables orders of magnitude higher $\Cs$ polarizations than that obtained using thermally polarized P1 centers under comparable conditions~\cite{waddington17}. The polarization builds up in under 60s of optical pumping (\zfr{results}B) and points to the efficiency of the underlying DNP mechanism. The slow rise in polarization after initial exponential growth is a direct indication of spin diffusion in our system -- the $\Cs$'s close to the NV centers being highly polarized and spin diffusing their polarization to nuclei further away. Note that for clarity the signals in \zfr{results}A have their noise unit-normalized, and a single shot DNP signal has about 25 times the signal-to-noise (SNR) of the 7T thermal signal obtained after $\app$7hr of averaging -- a time gain of over 5 orders of magnitude for identical SNR. In fact, the $\Cs$ signal is so greatly enhanced that it enables detection of a \I{single} 200$\mu$m particle in a single shot with unit SNR.

Our technique allows simple control of the hyperpolarization direction (\zfr{results}C). Sweeping the microwaves in a ramped fashion from low-to-high frequency leads to nuclear polarization aligned to $\T{B}_{\R{pol}}$. Anti-alignment can be achieved accordingly by sweeping from high-to-low-frequency. This allows on-demand control of the sign of polarization. As expected, a triangular sweep pattern with equal amounts of high-to-low and low-to-high frequency sweeps leads to destructive interference in alternate periods, and no net polarization buildup (\zfr{results}C). This feature may prove useful, for example, for common-mode noise rejection in signal enhanced nanodiamond imaging. Such room temperature hyperpolarized MRI would provide a complimentary, non-invasive, 3D imaging modality to high NV density nanodiamonds presently utilized as non-blinking fluorescent biomarkers~\cite{yu05,Chang08,Bumb13}. Through our method, functionalized diamond microparticles in solution can be hyperpolarized with modest optical power ($\sim$ 1mW/200$\mu$m particle) and single shot detection sensitivity (\zfr{particle_scaling}). This will open up new possibilities for targeted bio-sensing~\cite{liu07,fu07}.  The use of NV centers instead of persistent radicals as a source of electron spin in DNP will eliminate
potential undesirable effects in in-vivo imaging such as oxidative stress
without proper filtration. 

Using the notation in the energy diagram of \zfr{mechanism}A, the process of nuclear spin hyperpolarization can be better understood in the rotating frame, where resonances take the form of avoided crossings. As one traverses the full set of $m_S=0\leftrightarrow m_S=\pm1$ transitions, moderately fast sweep rates make the more-weakly avoided crossings partially non-adiabatic, thus resulting in a selective population transfer between the different branches and consequently the generation of net nuclear spin polarization. As an illustration, consider the case of a positive hyperfine coupling ($A_{zz} = +0.5$ MHz) shown in \zfr{mechanism}B in the subset $m_S=0\leftrightarrow m_S=-1$. Assuming, for simplicity, the NV spin is in the $m_S=0$ state, nuclear spins polarize positively as one sweeps the Landau-Zener crossing from low to high frequencies; similarly, a negative polarization arises if one starts from the right side of the crossing and the direction of the sweep is reversed (not shown for brevity). Central to this proposed polarization process are the differential Landau-Zener transition probabilities, selectively favoring in this case the transfer of populations between branches with different electron and nuclear spin quantum numbers. The resulting nuclear spin polarization is negligible if the population transfer throughout the Landau-Zener crossings is complete (the fully non-adiabatic limit) meaning that the optimum is attained at some intermediate sweep rate, consistent with our observations (\zfr{results}D). 

The dynamics for negative hyperfine couplings are qualitatively different. The more-weakly avoided crossings occur between branches within the same electron spin manifold with the consequence that the nuclear spin polarization buildup becomes inefficient in either sweep direction (\zfr{mechanism}C). In other words, as one sweeps the set of transitions connecting the $m_S=0$ and $m_S=-1$ manifolds, only carbon spins with positive hyperfine couplings contribute to pump nuclear spin polarization. The converse is true for the $m_S=0\leftrightarrow m_S=+1$ subset of transitions (not shown), because, when $|A_{zz}|$ is greater than the nuclear Larmor frequency,the physics remains unchanged if we simultaneously reverse the signs of the electron spin projection number and hyperfine coupling constant. Since the number of nuclear spins experiencing positive and negative hyperfine couplings is comparable, it follows that the $\Cs$ signals from the $m_S=0\leftrightarrow m_S=+1$ manifold should feature similar amplitudes and the same (sweep-direction-dependent) sign as observed within the $m_S=0\leftrightarrow m_S=-1$ subset (\zfr{mechanism}B and \zfr{mechanism}C), which we confirm experimentally (see below). Implicit in the above picture is the assumption that the probability of optically exciting the NV spin during the Landau-Zener crossings is sufficiently low, a condition fulfilled herein given the relatively fast sweep rates ($\sim$40 MHz/ms at the optimum, \zfr{results}D) and low illumination intensities ($\sim$10 mW/mm\textsuperscript{2}) used in our experiments. 

As a crude approximation to the results in \zfr{results}D, we write the nuclear spin polarization $P \propto g(\dot{\omega}) \ q(\dot{\omega}) [1-Q(\dot{\omega})]$, where $Q(\dot{\omega})= \exp⁡(-\Omega^{2}/\dot{\omega})$ is the transition probability between branches differing only in the electron spin number, $\dot{\omega}$ is the frequency sweep rate, and $\Omega$ is the Rabi field amplitude. On the other hand, we express the transition probability between branches with different electron and nuclear spin numbers as $q(\dot{\omega}) = \exp(-\Lambda^{2}/\dot{\omega}) [1-Q(\dot{\omega})]$, where $\Lambda$ (in general, a function of the hyperfine coupling and magnetic field orientation) captures the effect of the smaller gap size near the narrower crossing, and the last factor ensures we regain the correct limit for fast sweeps (where $Q\gg q$ and the populations in each state before and after the crossing remain unchanged). Finally, $g(\dot{\omega}) = [1 - \exp(-\dot{\omega}/k)]$ with $k$ a fitting parameter, takes into account the cumulative effect of varying multiple sweeps within a fixed measurement time per point. The good agreement we attain with the experimental data set (solid trace in the main plot of \zfr{results}D) may be partly fortuitous, as the correct response must arise from an integral over all hyperfine couplings and magnetic field orientations, a task we will carry out in subsequent work.

Our preliminary calculations suggest that weakly-coupled carbons (i.e., $|A_{zz}|\lesssim$ 2 MHz) are dominant in driving the nuclear spin polarization process since they polarize efficiently and are comparatively more numerous than those in the first or second shell around the NV. Further, the frequency mismatch with bulk carbons (arising from second-order hyperfine contributions within the $m_S=0$ manifold, see Supplementary Material) is considerably lower for more-weakly coupled carbons, thus facilitating spin diffusion to the bulk. We caution, however, the proposed polarization mechanism should not be understood as exclusive, as other polarization channels involving more strongly coupled carbons ($|A_{zz}|\gtrsim$10 MHz) may also play a role, particularly in $\Cs$-enriched samples where the two-spin model used herein breaks down. 

\begin{figure}[t]
  \centering
  {\includegraphics[width=0.49\textwidth]{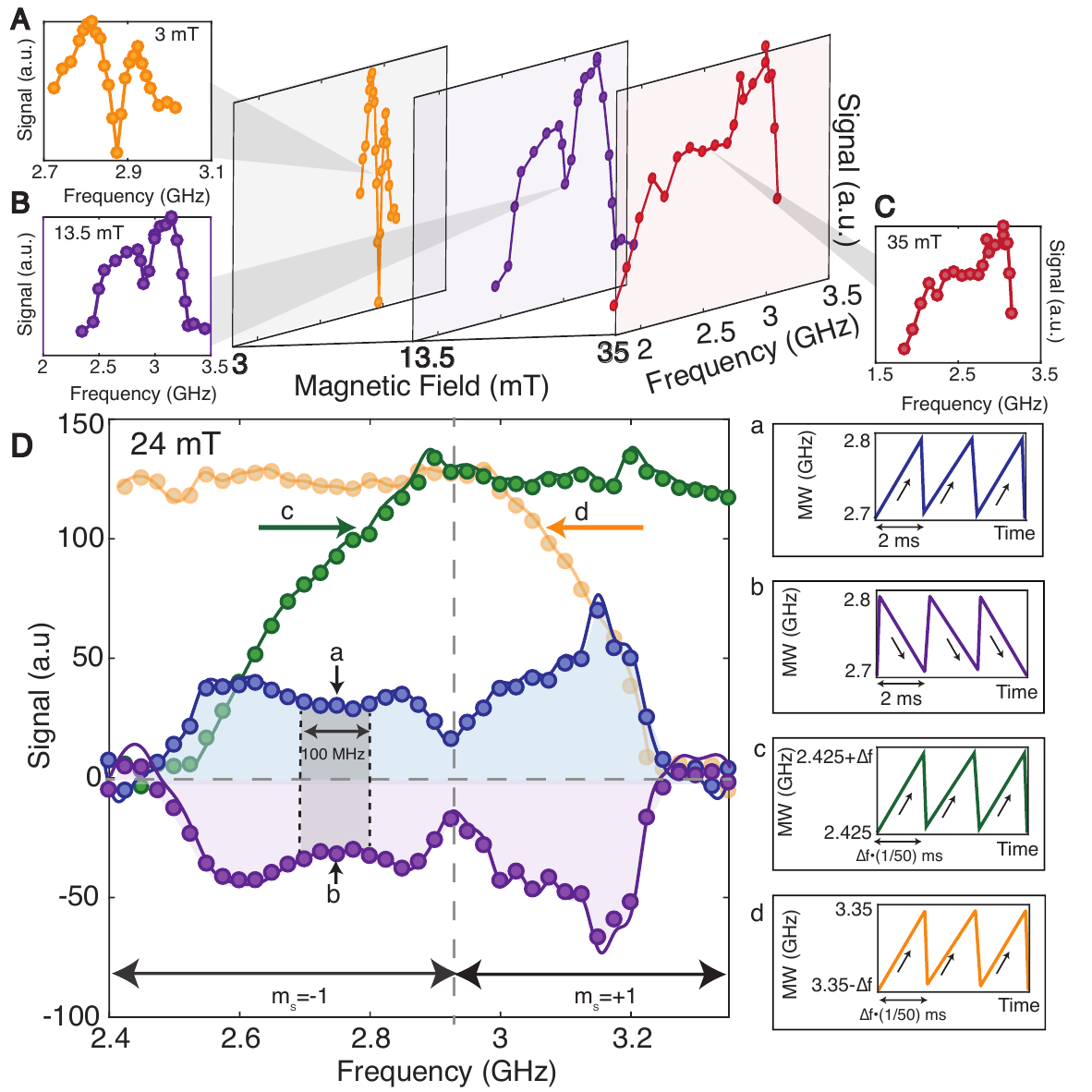}}
  \caption{\textbf{Contributions of different NV orientations to DNP.}  (A-C) \I{Electronic powder pattern mapped} by performing $\Cs$ DNP in a 100MHz window which is swept across in frequency space. This reports on the contributions of each window to the resulting signal and different orientations' relative contribution to DNP at different magnetic fields. Note that amplifier bandwidth limitations lead to an artificial cutoff at $\app$ 3.2GHz. (D) \I{Sign contributions from different NV orientations.} (a,b) Every part of the powder pattern, even if corresponding to different ($m_s=\pm 1$) electronic states, produces the same sign of hyperpolarization (shaded regions) that only depends on the direction of MW sweep. Solid lines are smoothened curves. Keeping the (c) lower (d) upper frequency of the DNP window fixed provides the cumulative contribution of different parts of the electronic spectrum to the polarization buildup. It shows that half the powder pattern is sufficient to saturate the polarization enhancement. Note that the we maintain the same differential sweep rate per unit spectral width equivalent to 40MHz/ms in all experiments. }
 \zfl{powder}	
\end{figure}

\begin{figure}[t]
  \centering
  {\includegraphics[width=0.45\textwidth]{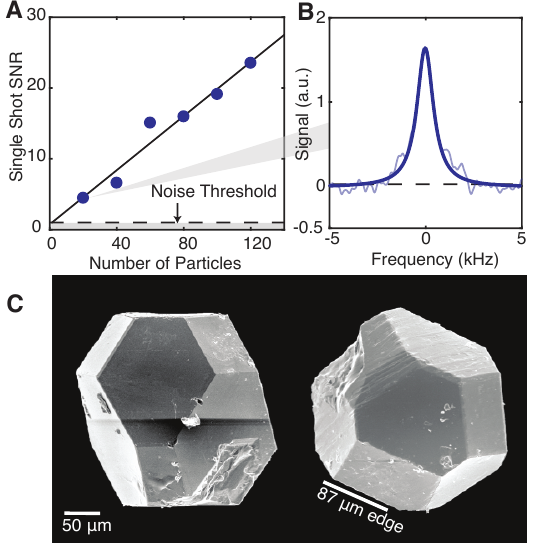}}
  \caption{\textbf{Minimum number of particles detectable.} (A) Panel indicates the single shot SNR scaling with the number of e6 200$\mu$m diamond microparticles in the sample tube. The experiments are performed by careful particle counting and averaging over 30 single-shot measurements for each accumulated collection of particles. By extrapolation, we determine that it is possible to obtain signal from a single particle with a single shot SNR of 0.994. (B) Average of 30 single-shot signals from 20 particles. (C) SEM micrograph (Hitachi S5000) of individual e6 HPHT diamond particles used in majority of the experiments. The particles have a uniform size distribution (edge length $87\pm3.9\ \mu$m), and a truncated octahedral shape set by particle growth conditions.}
 \zfl{particle_scaling}	
\end{figure}

To demonstrate more precisely how all NV center orientations contribute to the obtained hyperpolarization signal, we map the underlying electronic powder pattern via the $\Cs$ signal (\zfr{powder}). The DNP is performed by sweeping microwaves over a 100 MHz window which is subsequently moved in frequency space. While natively a convolution, the obtained signal does faithfully report on the electronic lineshape broadening as expected with increasing field (\zfr{powder}A-C). The results in \zfr{powder}D also exhibit another surprising aspect: unlike in conventional DNP methods e.g. solid, cross effects and thermal mixing~\cite{Hovav10}, where one expects a dispersion-like frequency dependence where certain parts of the electronic spectrum contribute positively or negatively to the enhancement, here \I{all} parts of the spectrum provide the same enhancement sign. This is also independent of whether one accesses the part of the spectrum corresponding to $m_s=+1$ or $m_s=-1$ electronic spin states. The hyperpolarization sign, as in \zfr{results}C, only depends on the direction of MW sweep. \zfr{powder}D also illustrates that only \I{half} the full electronic spectrum is sufficient to saturate the full extent of polarization, consistent with the fact that every NV center orientation is represented on either half of the powder pattern.

Finally, we measured the hyperpolarized signal from a varying number of particles (see \zfr{particle_scaling}) to estimate the minimum number of particles that can be detected. Single shot SNR averaged over 30 runs of the experiments were taken. \zfr{particle_scaling} demonstrates that after hyperpolarization with 40 seconds of laser pumping, it is possible to measure 0.994 particles (i.e. one \I{single} 200$\mu$m particle) with SNR$=1$ in one shot at 7T. This level of sensitivity has important implications in hyperpolarized imaging experiments, where the ability to detect a single particle would allow the hyperpolarized particles to act as MRI tracers. Indeed nanodiamonds have been shown to have long $\Cs$ $T_1$'s, even as large as 30 minutes at high fields~\cite{Rej15}, making them compelling for use as reporters in MRI.

\section*{Discussion}

These results demonstrate that in addition to providing large signal enhancements, our technique exploits a unique low-field DNP mechanism that is qualitatively different from others in the literature. We exploit the fact that the NV electrons are optically polarized at \I{any} field, and low fields mitigate their strong orientational dependence. From a technological standpoint, our technique is relatively simple to implement: hyperpolarization occurs at room temperature, MW amplifiers and sweep sources in the 2-4GHz range are low-cost and readily available, a simple stub antenna serves for MW irradiation, the laser and MW powers employed are very modest, and there is no requirement for magnetic field alignment. This opens up the possibility of constructing low-cost tabletop nanodiamond polarizers. Low field also comes with the added benefit of long target nuclear $T_1$s in the external nuclei due to reduced chemical shift anisotropy, allowing the potential for higher buildup of polarization. For instance, $\Cs$ spins in pyruvate, an important molecule in the metabolic cycle and cancer detection, can exceed 55s at 10mT~\cite{Chattergoon13}. When mildly frozen, for instance at liquid nitrogen temperature, the resulting $T_1$ can be nearly an hour~\cite{van11}. One limitation however, is the lower nuclear $T_1$ times in diamond at low fields, which limits the time period for spin diffusion within each particle. There is a strong indication that the $\Cs$ lifetime is set by their interactions to the dominant dipole coupled electronic spin bath consisting of nitrogen impurities (P1 centers).  Recent advances in diamond growth with high ($>$20\%) NV center conversion efficiency give optimism that they can be effectively mitigated~\cite{kucsko16}. Moreover, there is strong evidence~\cite{Panich15} that $\Cs$ lifetimes can be sufficiently long even for particles sizes down to 100nm.

In conclusion, we have developed a new DNP technique for polarization transfer from NV centers in diamond that is completely orientation independent, and have demonstrated its application for hyperpolarizing $\Cs$ nuclei in diamond microparticles to attain bulk polarization in excess of 0.25\%. The method was also found to work on smaller particles (1$\mu$m), although DNP enhancements were reduced on account of lower NV concentrations and shorter $T_1$. Our low-field optical DNP mechanism is unique in that the entire electronic spectrum contributes constructively to the polarization buildup with  on-demand control on the hyperpolarization direction.  Our work paves the way towards exploiting the large surface area intrinsic to diamond nanoparticles for optically hyperpolarizing a liquid at room temperature. Considering a spin diffusion constant of $D=8 \zt$ 10$^3\angstrom^2$/s~\cite{Van06}, one can potentially polarize $\sim$ 244$\mu$L of liquid per milligram of 100nm hyperpolarized diamonds in 125 seconds.  Moreover, the use of $\Cs$ enrichment in the diamond particles or as surface coatings will greatly enhance these spin diffusion rates to external liquids. We also envision utility of our method for signal enhancements in nanoscale MRI experiments mediated by NV sensors~\cite{Ajoy15}. Moreover, it presents an advance towards magnetic resonance imaging modalities for biosensing constructed out of optically hyperpolarized, surface functionalized, diamond particles. The diamond particles are non-toxic and can be functionalized ~\cite{Bumb13}, and hence in an MRI modality could potentially serve as a sensitive biosensor for disease detection~\cite{Spence01}, and in non-invasively characterizing fluid flow in bioreactors~\cite{Bouchard08}. 

\section*{Materials and Methods}
\zsl{materials}

\subsection*{Diamond Particles}

The experiments in this work were performed with $7.50\pm0.25$ mg ($287\pm27$ particles) of $200-250\mu$m diamond micro-particles from Element6, with $\app$1ppm NV center concentration. Scanning electron microscope (SEM) images (see \zfr{particle_scaling}C) show that the particles have a truncated octahedral shape though with some imperfections and irregularities. Overall the particles in the sample measured face-to-face from respective opposite faces are $200-250\mu$m in size, and approximately $400\mu$m measured diagonally edge-to-edge. Using the known scale of the images, all visible edge lengths were measured and the average edge length found to be $87\pm3.9\mu$m (marked in \zfr{particle_scaling}C). This average edge length was used for all further calculations of surface area and volume (see Supplemental Information).

\subsection*{Experimental Design}
 
A fast field cycling device is utilized to leverage rapid mechanical sample shuttling from the low field polarizing volume (1-30mT) composed of stress annealed iron (NETIC S3-6 alloy 0.062'' thick, Magnetic Shield Corp) to a widebore 7T superconducting magnet for detection. A conveyor belt actuator stage (Parker HMRB08) carries a rigidly fastened NMR tube (8mm) containing the diamond particles along the fringing field of the magnet to magnetic field extremes. A home-built NMR probe with a hollow opening was designed to allow rapid mechanical transfer of sample between the low field volume and the 7T magnet. The sample is channeled through a series of funnel-shaped guiding stages made of soft teflon that dynamically align the sample concentric to the magnetic bore to within 1mdeg. Shuttling a sample from 8mT to 7T takes $648\pm $2.5ms with high positional precision ($50\mu$m) at a maximum speed of 2m/s and acceleration of 30m/s$^{2}$. Motion of the actuator stage is synchronized to trigger at the end of a polarization cycle with inductive detection coinciding with the end of shuttling. NMR measurements were performed with a custom printed saddle coil tuned to 75.03MHz for $\Cs$ detection (see Supplemental Information).

\subsection*{Hyperpolarization of diamond}
Hyperpolarization is performed at the low field volume (1-30mT) for times up to a minute, after which the sample is rapidly shuttled to 7T for bulk  $\Cs$ measurement. Low field DNP is implemented by continuous irradiation of laser light that polarizes NV centers accompanied by frequency sweeps over the NV center powder pattern.  A Coherent Verdi laser (532 nm) delivers a continuous collimated beam that is expanded to match the diameter of the NMR tube containing the sample (8mm). The laser power was selected to conditions that maximize DNP transfer efficient, at a total laser power of 200mW over a 8mm beam diameter. (see Supplemental Information). Frequency sweeps are generated via voltage controlled oscillators (VCOs) (Minicircuits ZX95-3800A+, 1.9-3.7GHz). The VCOs are regulated by programmable ramp inputs to control the direction, bandwidth, and sweep rate of frequency sweeps. The outputs are combined and amplified with a 100W amplifier (Empower SKU-1146) before being directed to the sample via a stub-loop antenna (see Supplemental Information).


\I{Author contributions:} -- A.A. proposed the method of low field DNP with frequency sweeps, and designed the experimental setup and protocols. A.A.,E.D.,R.N,X.L and B.S. built the polarization setup and fast-field cycler. A.A, K.L, R.N, G.W and X.L. performed experiments. A.A., D.A., G.L., A.L, P.R. analyzed data and wrote software. P.Z., A.A., S.D. and C.A.M. performed theoretical simulations. D.S, C.A.M, J.A.R and D.P. advised on several aspects of theory and experiments. A.A. wrote the paper with inputs from all authors. All authors reviewed the manuscript and suggested improvements. A.P. supervised the overall research effort.

\I{Acknowledgments:} -- C.A.M. acknowledges support from the National Science Foundation through grants NSF-1309640 and NSF-1401632, and from Research Corporation for Science Advancement through a FRED Award, and access to the facilities and research infrastructure of the NSF CREST Center IDEALS, grant number NSF-HRD-1547830. All data needed to evaluate the conclusions in the paper are present in the paper and/or the Supplementary Materials. Additional data related to this paper may be requested from the authors. All correspondence and request for materials should be addressed to A.A. (ashokaj@berkeley.edu).

\bibliography{Biblio}

\begin{thebibliography}{40}%
\makeatletter
\providecommand \@ifxundefined [1]{%
 \@ifx{#1\undefined}
}%
\providecommand \@ifnum [1]{%
 \ifnum #1\expandafter \@firstoftwo
 \else \expandafter \@secondoftwo
 \fi
}%
\providecommand \@ifx [1]{%
 \ifx #1\expandafter \@firstoftwo
 \else \expandafter \@secondoftwo
 \fi
}%
\providecommand \natexlab [1]{#1}%
\providecommand \enquote  [1]{``#1''}%
\providecommand \bibnamefont  [1]{#1}%
\providecommand \bibfnamefont [1]{#1}%
\providecommand \citenamefont [1]{#1}%
\providecommand \href@noop [0]{\@secondoftwo}%
\providecommand \href [0]{\begingroup \@sanitize@url \@href}%
\providecommand \@href[1]{\@@startlink{#1}\@@href}%
\providecommand \@@href[1]{\endgroup#1\@@endlink}%
\providecommand \@sanitize@url [0]{\catcode `\\12\catcode `\$12\catcode
  `\&12\catcode `\#12\catcode `\^12\catcode `\_12\catcode `\%12\relax}%
\providecommand \@@startlink[1]{}%
\providecommand \@@endlink[0]{}%
\providecommand \url  [0]{\begingroup\@sanitize@url \@url }%
\providecommand \@url [1]{\endgroup\@href {#1}{\urlprefix }}%
\providecommand \urlprefix  [0]{URL }%
\providecommand \Eprint [0]{\href }%
\providecommand \doibase [0]{http://dx.doi.org/}%
\providecommand \selectlanguage [0]{\@gobble}%
\providecommand \bibinfo  [0]{\@secondoftwo}%
\providecommand \bibfield  [0]{\@secondoftwo}%
\providecommand \translation [1]{[#1]}%
\providecommand \BibitemOpen [0]{}%
\providecommand \bibitemStop [0]{}%
\providecommand \bibitemNoStop [0]{.\EOS\space}%
\providecommand \EOS [0]{\spacefactor3000\relax}%
\providecommand \BibitemShut  [1]{\csname bibitem#1\endcsname}%
\let\auto@bib@innerbib\@empty
\bibitem [{\citenamefont {Ernst}\ \emph {et~al.}(1987)\citenamefont {Ernst},
  \citenamefont {Bodenhausen},\ and\ \citenamefont {Wokaun}}]{Ernst}%
  \BibitemOpen
  \bibfield  {author} {\bibinfo {author} {\bibfnamefont {R.}~\bibnamefont
  {Ernst}}, \bibinfo {author} {\bibfnamefont {G.}~\bibnamefont {Bodenhausen}},
  \ and\ \bibinfo {author} {\bibfnamefont {A.}~\bibnamefont {Wokaun}},\
  }\href@noop {} {\emph {\bibinfo {title} {Principles of nuclear magnetic
  resonance in one and two dimensions}}}\ (\bibinfo  {publisher} {Clarendon
  Press Oxford},\ \bibinfo {year} {1987})\BibitemShut {NoStop}%
\bibitem [{\citenamefont {W{\"u}thrich}(2003)}]{Wuthrich03}%
  \BibitemOpen
  \bibfield  {author} {\bibinfo {author} {\bibfnamefont {K.}~\bibnamefont
  {W{\"u}thrich}},\ }\href@noop {} {\bibfield  {journal} {\bibinfo  {journal}
  {Angewandte Chemie International Edition}\ }\textbf {\bibinfo {volume}
  {42}},\ \bibinfo {pages} {3340} (\bibinfo {year} {2003})}\BibitemShut
  {NoStop}%
\bibitem [{\citenamefont {Abragam}\ and\ \citenamefont
  {Goldman}(1978)}]{Abragam78}%
  \BibitemOpen
  \bibfield  {author} {\bibinfo {author} {\bibfnamefont {A.}~\bibnamefont
  {Abragam}}\ and\ \bibinfo {author} {\bibfnamefont {M.}~\bibnamefont
  {Goldman}},\ }\href@noop {} {\bibfield  {journal} {\bibinfo  {journal}
  {Reports on Progress in Physics}\ }\textbf {\bibinfo {volume} {41}},\
  \bibinfo {pages} {395} (\bibinfo {year} {1978})}\BibitemShut {NoStop}%
\bibitem [{\citenamefont {Ardenkjaer-Larsen}\ \emph {et~al.}(2003)\citenamefont
  {Ardenkjaer-Larsen}, \citenamefont {Fridlund}, \citenamefont {Gram},
  \citenamefont {Hansson}, \citenamefont {Hansson}, \citenamefont {Lerche},
  \citenamefont {Servin}, \citenamefont {Thaning},\ and\ \citenamefont
  {Golman}}]{ArdenkjaerLarsen03}%
  \BibitemOpen
  \bibfield  {author} {\bibinfo {author} {\bibfnamefont {J.~H.}\ \bibnamefont
  {Ardenkjaer-Larsen}}, \bibinfo {author} {\bibfnamefont {B.}~\bibnamefont
  {Fridlund}}, \bibinfo {author} {\bibfnamefont {A.}~\bibnamefont {Gram}},
  \bibinfo {author} {\bibfnamefont {G.}~\bibnamefont {Hansson}}, \bibinfo
  {author} {\bibfnamefont {L.}~\bibnamefont {Hansson}}, \bibinfo {author}
  {\bibfnamefont {M.~H.}\ \bibnamefont {Lerche}}, \bibinfo {author}
  {\bibfnamefont {R.}~\bibnamefont {Servin}}, \bibinfo {author} {\bibfnamefont
  {M.}~\bibnamefont {Thaning}}, \ and\ \bibinfo {author} {\bibfnamefont
  {K.}~\bibnamefont {Golman}},\ }\href {\doibase 10.1073/pnas.1733835100}
  {\bibfield  {journal} {\bibinfo  {journal} {Proc. Nat. Acad. Sc.}\ }\textbf
  {\bibinfo {volume} {100}},\ \bibinfo {pages} {10158} (\bibinfo {year}
  {2003})}\BibitemShut {NoStop}%
\bibitem [{\citenamefont {Maly}\ \emph {et~al.}(2008)\citenamefont {Maly},
  \citenamefont {Debelouchina}, \citenamefont {Bajaj}, \citenamefont {Hu},
  \citenamefont {Joo}, \citenamefont {MakJurkauskas}, \citenamefont {Sirigiri},
  \citenamefont {van~der Wel}, \citenamefont {Herzfeld}, \citenamefont
  {Temkin},\ and\ \citenamefont {Griffin}}]{Maly08}%
  \BibitemOpen
  \bibfield  {author} {\bibinfo {author} {\bibfnamefont {T.}~\bibnamefont
  {Maly}}, \bibinfo {author} {\bibfnamefont {G.~T.}\ \bibnamefont
  {Debelouchina}}, \bibinfo {author} {\bibfnamefont {V.~S.}\ \bibnamefont
  {Bajaj}}, \bibinfo {author} {\bibfnamefont {K.-N.}\ \bibnamefont {Hu}},
  \bibinfo {author} {\bibfnamefont {C.-G.}\ \bibnamefont {Joo}}, \bibinfo
  {author} {\bibfnamefont {M.~L.}\ \bibnamefont {MakJurkauskas}}, \bibinfo
  {author} {\bibfnamefont {J.~R.}\ \bibnamefont {Sirigiri}}, \bibinfo {author}
  {\bibfnamefont {P.~C.~A.}\ \bibnamefont {van~der Wel}}, \bibinfo {author}
  {\bibfnamefont {J.}~\bibnamefont {Herzfeld}}, \bibinfo {author}
  {\bibfnamefont {R.~J.}\ \bibnamefont {Temkin}}, \ and\ \bibinfo {author}
  {\bibfnamefont {R.~G.}\ \bibnamefont {Griffin}},\ }\href {\doibase
  10.1063/1.2833582} {\bibfield  {journal} {\bibinfo  {journal} {The Journal of
  Chemical Physics}\ }\textbf {\bibinfo {volume} {128}},\ \bibinfo {eid}
  {052211} (\bibinfo {year} {2008})}\BibitemShut {NoStop}%
\bibitem [{\citenamefont {Abrams}\ \emph {et~al.}(2014)\citenamefont {Abrams},
  \citenamefont {Trusheim}, \citenamefont {Englund}, \citenamefont {Shattuck},\
  and\ \citenamefont {Meriles}}]{Abrams14}%
  \BibitemOpen
  \bibfield  {author} {\bibinfo {author} {\bibfnamefont {D.}~\bibnamefont
  {Abrams}}, \bibinfo {author} {\bibfnamefont {M.~E.}\ \bibnamefont
  {Trusheim}}, \bibinfo {author} {\bibfnamefont {D.~R.}\ \bibnamefont
  {Englund}}, \bibinfo {author} {\bibfnamefont {M.~D.}\ \bibnamefont
  {Shattuck}}, \ and\ \bibinfo {author} {\bibfnamefont {C.~A.}\ \bibnamefont
  {Meriles}},\ }\href@noop {} {\bibfield  {journal} {\bibinfo  {journal} {Nano
  letters}\ }\textbf {\bibinfo {volume} {14}},\ \bibinfo {pages} {2471}
  (\bibinfo {year} {2014})}\BibitemShut {NoStop}%
\bibitem [{\citenamefont {Jelezko}\ and\ \citenamefont
  {Wrachtrup}(2006)}]{Jelezko06}%
  \BibitemOpen
  \bibfield  {author} {\bibinfo {author} {\bibfnamefont {F.}~\bibnamefont
  {Jelezko}}\ and\ \bibinfo {author} {\bibfnamefont {J.}~\bibnamefont
  {Wrachtrup}},\ }\href {\doibase 10.1002/pssa.200671403} {\bibfield  {journal}
  {\bibinfo  {journal} {Physica Status Solidi (A)}\ }\textbf {\bibinfo {volume}
  {203}},\ \bibinfo {pages} {3207} (\bibinfo {year} {2006})}\BibitemShut
  {NoStop}%
\bibitem [{\citenamefont {Balasubramanian}\ \emph {et~al.}(2009)\citenamefont
  {Balasubramanian}, \citenamefont {Neumann}, \citenamefont {Twitchen},
  \citenamefont {Markham}, \citenamefont {Kolesov}, \citenamefont {Mizuochi},
  \citenamefont {Isoya}, \citenamefont {Achard}, \citenamefont {Beck},
  \citenamefont {Tissler}, \citenamefont {Jacques}, \citenamefont {Hemmer},
  \citenamefont {Jelezko},\ and\ \citenamefont
  {Wrachtrup}}]{Balasubramanian09}%
  \BibitemOpen
  \bibfield  {author} {\bibinfo {author} {\bibfnamefont {G.}~\bibnamefont
  {Balasubramanian}}, \bibinfo {author} {\bibfnamefont {P.}~\bibnamefont
  {Neumann}}, \bibinfo {author} {\bibfnamefont {D.}~\bibnamefont {Twitchen}},
  \bibinfo {author} {\bibfnamefont {M.}~\bibnamefont {Markham}}, \bibinfo
  {author} {\bibfnamefont {R.}~\bibnamefont {Kolesov}}, \bibinfo {author}
  {\bibfnamefont {N.}~\bibnamefont {Mizuochi}}, \bibinfo {author}
  {\bibfnamefont {J.}~\bibnamefont {Isoya}}, \bibinfo {author} {\bibfnamefont
  {J.}~\bibnamefont {Achard}}, \bibinfo {author} {\bibfnamefont
  {J.}~\bibnamefont {Beck}}, \bibinfo {author} {\bibfnamefont {J.}~\bibnamefont
  {Tissler}}, \bibinfo {author} {\bibfnamefont {V.}~\bibnamefont {Jacques}},
  \bibinfo {author} {\bibfnamefont {P.~R.}\ \bibnamefont {Hemmer}}, \bibinfo
  {author} {\bibfnamefont {F.}~\bibnamefont {Jelezko}}, \ and\ \bibinfo
  {author} {\bibfnamefont {J.}~\bibnamefont {Wrachtrup}},\ }\href {\doibase
  10.1038/nmat2420} {\bibfield  {journal} {\bibinfo  {journal} {Nat Mater}\
  }\textbf {\bibinfo {volume} {8}},\ \bibinfo {pages} {383} (\bibinfo {year}
  {2009})}\BibitemShut {NoStop}%
\bibitem [{\citenamefont {Staudacher}\ \emph {et~al.}(2013)\citenamefont
  {Staudacher}, \citenamefont {Shi}, \citenamefont {Pezzagna}, \citenamefont
  {Meijer}, \citenamefont {Du}, \citenamefont {Meriles}, \citenamefont
  {Reinhard},\ and\ \citenamefont {Wrachtrup}}]{Staudacher13}%
  \BibitemOpen
  \bibfield  {author} {\bibinfo {author} {\bibfnamefont {T.}~\bibnamefont
  {Staudacher}}, \bibinfo {author} {\bibfnamefont {F.}~\bibnamefont {Shi}},
  \bibinfo {author} {\bibfnamefont {S.}~\bibnamefont {Pezzagna}}, \bibinfo
  {author} {\bibfnamefont {J.}~\bibnamefont {Meijer}}, \bibinfo {author}
  {\bibfnamefont {J.}~\bibnamefont {Du}}, \bibinfo {author} {\bibfnamefont
  {C.~A.}\ \bibnamefont {Meriles}}, \bibinfo {author} {\bibfnamefont
  {F.}~\bibnamefont {Reinhard}}, \ and\ \bibinfo {author} {\bibfnamefont
  {J.}~\bibnamefont {Wrachtrup}},\ }\href {\doibase 10.1126/science.1231675}
  {\bibfield  {journal} {\bibinfo  {journal} {Science}\ }\textbf {\bibinfo
  {volume} {339}},\ \bibinfo {pages} {561} (\bibinfo {year}
  {2013})}\BibitemShut {NoStop}%
\bibitem [{\citenamefont {Lovchinsky}\ \emph {et~al.}(2016)\citenamefont
  {Lovchinsky}, \citenamefont {Sushkov}, \citenamefont {Urbach}, \citenamefont
  {de~Leon}, \citenamefont {Choi}, \citenamefont {De~Greve}, \citenamefont
  {Evans}, \citenamefont {Gertner}, \citenamefont {Bersin}, \citenamefont
  {M{\"u}ller} \emph {et~al.}}]{Lovchinsky16}%
  \BibitemOpen
  \bibfield  {author} {\bibinfo {author} {\bibfnamefont {I.}~\bibnamefont
  {Lovchinsky}}, \bibinfo {author} {\bibfnamefont {A.}~\bibnamefont {Sushkov}},
  \bibinfo {author} {\bibfnamefont {E.}~\bibnamefont {Urbach}}, \bibinfo
  {author} {\bibfnamefont {N.}~\bibnamefont {de~Leon}}, \bibinfo {author}
  {\bibfnamefont {S.}~\bibnamefont {Choi}}, \bibinfo {author} {\bibfnamefont
  {K.}~\bibnamefont {De~Greve}}, \bibinfo {author} {\bibfnamefont
  {R.}~\bibnamefont {Evans}}, \bibinfo {author} {\bibfnamefont
  {R.}~\bibnamefont {Gertner}}, \bibinfo {author} {\bibfnamefont
  {E.}~\bibnamefont {Bersin}}, \bibinfo {author} {\bibfnamefont
  {C.}~\bibnamefont {M{\"u}ller}},  \emph {et~al.},\ }\href@noop {} {\bibfield
  {journal} {\bibinfo  {journal} {Science}\ }\textbf {\bibinfo {volume}
  {351}},\ \bibinfo {pages} {836} (\bibinfo {year} {2016})}\BibitemShut
  {NoStop}%
\bibitem [{\citenamefont {Fischer}\ \emph {et~al.}(2013)\citenamefont
  {Fischer}, \citenamefont {Bretschneider}, \citenamefont {London},
  \citenamefont {Budker}, \citenamefont {Gershoni},\ and\ \citenamefont
  {Frydman}}]{Fischer13}%
  \BibitemOpen
  \bibfield  {author} {\bibinfo {author} {\bibfnamefont {R.}~\bibnamefont
  {Fischer}}, \bibinfo {author} {\bibfnamefont {C.~O.}\ \bibnamefont
  {Bretschneider}}, \bibinfo {author} {\bibfnamefont {P.}~\bibnamefont
  {London}}, \bibinfo {author} {\bibfnamefont {D.}~\bibnamefont {Budker}},
  \bibinfo {author} {\bibfnamefont {D.}~\bibnamefont {Gershoni}}, \ and\
  \bibinfo {author} {\bibfnamefont {L.}~\bibnamefont {Frydman}},\ }\href@noop
  {} {\bibfield  {journal} {\bibinfo  {journal} {Physical review letters}\
  }\textbf {\bibinfo {volume} {111}},\ \bibinfo {pages} {057601} (\bibinfo
  {year} {2013})}\BibitemShut {NoStop}%
\bibitem [{\citenamefont {{\'A}lvarez}\ \emph {et~al.}(2015)\citenamefont
  {{\'A}lvarez}, \citenamefont {Bretschneider}, \citenamefont {Fischer},
  \citenamefont {London}, \citenamefont {Kanda}, \citenamefont {Onoda},
  \citenamefont {Isoya}, \citenamefont {Gershoni},\ and\ \citenamefont
  {Frydman}}]{Alvarez15}%
  \BibitemOpen
  \bibfield  {author} {\bibinfo {author} {\bibfnamefont {G.~A.}\ \bibnamefont
  {{\'A}lvarez}}, \bibinfo {author} {\bibfnamefont {C.~O.}\ \bibnamefont
  {Bretschneider}}, \bibinfo {author} {\bibfnamefont {R.}~\bibnamefont
  {Fischer}}, \bibinfo {author} {\bibfnamefont {P.}~\bibnamefont {London}},
  \bibinfo {author} {\bibfnamefont {H.}~\bibnamefont {Kanda}}, \bibinfo
  {author} {\bibfnamefont {S.}~\bibnamefont {Onoda}}, \bibinfo {author}
  {\bibfnamefont {J.}~\bibnamefont {Isoya}}, \bibinfo {author} {\bibfnamefont
  {D.}~\bibnamefont {Gershoni}}, \ and\ \bibinfo {author} {\bibfnamefont
  {L.}~\bibnamefont {Frydman}},\ }\href@noop {} {\bibfield  {journal} {\bibinfo
   {journal} {Nature communications}\ }\textbf {\bibinfo {volume} {6}}
  (\bibinfo {year} {2015})}\BibitemShut {NoStop}%
\bibitem [{\citenamefont {King}\ \emph {et~al.}(2015)\citenamefont {King},
  \citenamefont {Jeong}, \citenamefont {Vassiliou}, \citenamefont {Shin},
  \citenamefont {Page}, \citenamefont {Avalos}, \citenamefont {Wang},\ and\
  \citenamefont {Pines}}]{King15}%
  \BibitemOpen
  \bibfield  {author} {\bibinfo {author} {\bibfnamefont {J.~P.}\ \bibnamefont
  {King}}, \bibinfo {author} {\bibfnamefont {K.}~\bibnamefont {Jeong}},
  \bibinfo {author} {\bibfnamefont {C.~C.}\ \bibnamefont {Vassiliou}}, \bibinfo
  {author} {\bibfnamefont {C.~S.}\ \bibnamefont {Shin}}, \bibinfo {author}
  {\bibfnamefont {R.~H.}\ \bibnamefont {Page}}, \bibinfo {author}
  {\bibfnamefont {C.~E.}\ \bibnamefont {Avalos}}, \bibinfo {author}
  {\bibfnamefont {H.-J.}\ \bibnamefont {Wang}}, \ and\ \bibinfo {author}
  {\bibfnamefont {A.}~\bibnamefont {Pines}},\ }\href@noop {} {\bibfield
  {journal} {\bibinfo  {journal} {Nature communications}\ }\textbf {\bibinfo
  {volume} {6}},\ \bibinfo {pages} {8965} (\bibinfo {year} {2015})}\BibitemShut
  {NoStop}%
\bibitem [{\citenamefont {Pagliero}\ \emph {et~al.}(2018)\citenamefont
  {Pagliero}, \citenamefont {Rao}, \citenamefont {Zangara}, \citenamefont
  {Dhomkar}, \citenamefont {Wong}, \citenamefont {Abril}, \citenamefont
  {Aslam}, \citenamefont {Parker}, \citenamefont {King}, \citenamefont
  {Avalos}, \citenamefont {Ajoy}, \citenamefont {Wrachtrup}, \citenamefont
  {Pines},\ and\ \citenamefont {Meriles}}]{Pagliero2018}%
  \BibitemOpen
  \bibfield  {author} {\bibinfo {author} {\bibfnamefont {D.}~\bibnamefont
  {Pagliero}}, \bibinfo {author} {\bibfnamefont {K.~R.~K.}\ \bibnamefont
  {Rao}}, \bibinfo {author} {\bibfnamefont {P.~R.}\ \bibnamefont {Zangara}},
  \bibinfo {author} {\bibfnamefont {S.}~\bibnamefont {Dhomkar}}, \bibinfo
  {author} {\bibfnamefont {H.~H.}\ \bibnamefont {Wong}}, \bibinfo {author}
  {\bibfnamefont {A.}~\bibnamefont {Abril}}, \bibinfo {author} {\bibfnamefont
  {N.}~\bibnamefont {Aslam}}, \bibinfo {author} {\bibfnamefont
  {A.}~\bibnamefont {Parker}}, \bibinfo {author} {\bibfnamefont
  {J.}~\bibnamefont {King}}, \bibinfo {author} {\bibfnamefont {C.~E.}\
  \bibnamefont {Avalos}}, \bibinfo {author} {\bibfnamefont {A.}~\bibnamefont
  {Ajoy}}, \bibinfo {author} {\bibfnamefont {J.}~\bibnamefont {Wrachtrup}},
  \bibinfo {author} {\bibfnamefont {A.}~\bibnamefont {Pines}}, \ and\ \bibinfo
  {author} {\bibfnamefont {C.~A.}\ \bibnamefont {Meriles}},\ }\href {\doibase
  10.1103/PhysRevB.97.024422} {\bibfield  {journal} {\bibinfo  {journal} {Phys.
  Rev. B}\ }\textbf {\bibinfo {volume} {97}},\ \bibinfo {pages} {024422}
  (\bibinfo {year} {2018})}\BibitemShut {NoStop}%
\bibitem [{\citenamefont {Chen}\ \emph {et~al.}(2015)\citenamefont {Chen},
  \citenamefont {Schwarz}, \citenamefont {Jelezko}, \citenamefont {Retzker},\
  and\ \citenamefont {Plenio}}]{chen15b}%
  \BibitemOpen
  \bibfield  {author} {\bibinfo {author} {\bibfnamefont {Q.}~\bibnamefont
  {Chen}}, \bibinfo {author} {\bibfnamefont {I.}~\bibnamefont {Schwarz}},
  \bibinfo {author} {\bibfnamefont {F.}~\bibnamefont {Jelezko}}, \bibinfo
  {author} {\bibfnamefont {A.}~\bibnamefont {Retzker}}, \ and\ \bibinfo
  {author} {\bibfnamefont {M.}~\bibnamefont {Plenio}},\ }\href@noop {}
  {\bibfield  {journal} {\bibinfo  {journal} {Physical Review B}\ }\textbf
  {\bibinfo {volume} {92}},\ \bibinfo {pages} {184420} (\bibinfo {year}
  {2015})}\BibitemShut {NoStop}%
\bibitem [{\citenamefont {Scott}\ \emph {et~al.}(2016)\citenamefont {Scott},
  \citenamefont {Drake},\ and\ \citenamefont {Reimer}}]{Scott16}%
  \BibitemOpen
  \bibfield  {author} {\bibinfo {author} {\bibfnamefont {E.}~\bibnamefont
  {Scott}}, \bibinfo {author} {\bibfnamefont {M.}~\bibnamefont {Drake}}, \ and\
  \bibinfo {author} {\bibfnamefont {J.~A.}\ \bibnamefont {Reimer}},\
  }\href@noop {} {\bibfield  {journal} {\bibinfo  {journal} {Journal of
  Magnetic Resonance}\ }\textbf {\bibinfo {volume} {264}},\ \bibinfo {pages}
  {154} (\bibinfo {year} {2016})}\BibitemShut {NoStop}%
\bibitem [{\citenamefont {London}\ \emph {et~al.}(2013)\citenamefont {London},
  \citenamefont {Scheuer}, \citenamefont {Cai}, \citenamefont {Schwarz},
  \citenamefont {Retzker}, \citenamefont {Plenio}, \citenamefont {Katagiri},
  \citenamefont {Teraji}, \citenamefont {Koizumi}, \citenamefont {Isoya} \emph
  {et~al.}}]{London13}%
  \BibitemOpen
  \bibfield  {author} {\bibinfo {author} {\bibfnamefont {P.}~\bibnamefont
  {London}}, \bibinfo {author} {\bibfnamefont {J.}~\bibnamefont {Scheuer}},
  \bibinfo {author} {\bibfnamefont {J.-M.}\ \bibnamefont {Cai}}, \bibinfo
  {author} {\bibfnamefont {I.}~\bibnamefont {Schwarz}}, \bibinfo {author}
  {\bibfnamefont {A.}~\bibnamefont {Retzker}}, \bibinfo {author} {\bibfnamefont
  {M.}~\bibnamefont {Plenio}}, \bibinfo {author} {\bibfnamefont
  {M.}~\bibnamefont {Katagiri}}, \bibinfo {author} {\bibfnamefont
  {T.}~\bibnamefont {Teraji}}, \bibinfo {author} {\bibfnamefont
  {S.}~\bibnamefont {Koizumi}}, \bibinfo {author} {\bibfnamefont
  {J.}~\bibnamefont {Isoya}},  \emph {et~al.},\ }\href@noop {} {\bibfield
  {journal} {\bibinfo  {journal} {Physical review letters}\ }\textbf {\bibinfo
  {volume} {111}},\ \bibinfo {pages} {067601} (\bibinfo {year}
  {2013})}\BibitemShut {NoStop}%
\bibitem [{\citenamefont {Scheuer}\ \emph {et~al.}(2016)\citenamefont
  {Scheuer}, \citenamefont {Schwartz}, \citenamefont {Chen}, \citenamefont
  {Schulze-S{\"u}nninghausen}, \citenamefont {Carl}, \citenamefont {H{\"o}fer},
  \citenamefont {Retzker}, \citenamefont {Sumiya}, \citenamefont {Isoya},
  \citenamefont {Luy} \emph {et~al.}}]{Scheuer16}%
  \BibitemOpen
  \bibfield  {author} {\bibinfo {author} {\bibfnamefont {J.}~\bibnamefont
  {Scheuer}}, \bibinfo {author} {\bibfnamefont {I.}~\bibnamefont {Schwartz}},
  \bibinfo {author} {\bibfnamefont {Q.}~\bibnamefont {Chen}}, \bibinfo {author}
  {\bibfnamefont {D.}~\bibnamefont {Schulze-S{\"u}nninghausen}}, \bibinfo
  {author} {\bibfnamefont {P.}~\bibnamefont {Carl}}, \bibinfo {author}
  {\bibfnamefont {P.}~\bibnamefont {H{\"o}fer}}, \bibinfo {author}
  {\bibfnamefont {A.}~\bibnamefont {Retzker}}, \bibinfo {author} {\bibfnamefont
  {H.}~\bibnamefont {Sumiya}}, \bibinfo {author} {\bibfnamefont
  {J.}~\bibnamefont {Isoya}}, \bibinfo {author} {\bibfnamefont
  {B.}~\bibnamefont {Luy}},  \emph {et~al.},\ }\href@noop {} {\bibfield
  {journal} {\bibinfo  {journal} {New Journal of Physics}\ }\textbf {\bibinfo
  {volume} {18}},\ \bibinfo {pages} {013040} (\bibinfo {year}
  {2016})}\BibitemShut {NoStop}%
\bibitem [{\citenamefont {Broadway}\ \emph {et~al.}(2017)\citenamefont
  {Broadway}, \citenamefont {Tetienne}, \citenamefont {Stacey}, \citenamefont
  {Wood}, \citenamefont {Simpson}, \citenamefont {Hall},\ and\ \citenamefont
  {Hollenberg}}]{Broadway17}%
  \BibitemOpen
  \bibfield  {author} {\bibinfo {author} {\bibfnamefont {D.~A.}\ \bibnamefont
  {Broadway}}, \bibinfo {author} {\bibfnamefont {J.-P.}\ \bibnamefont
  {Tetienne}}, \bibinfo {author} {\bibfnamefont {A.}~\bibnamefont {Stacey}},
  \bibinfo {author} {\bibfnamefont {J.~D.}\ \bibnamefont {Wood}}, \bibinfo
  {author} {\bibfnamefont {D.~A.}\ \bibnamefont {Simpson}}, \bibinfo {author}
  {\bibfnamefont {L.~T.}\ \bibnamefont {Hall}}, \ and\ \bibinfo {author}
  {\bibfnamefont {L.~C.}\ \bibnamefont {Hollenberg}},\ }\href@noop {}
  {\bibfield  {journal} {\bibinfo  {journal} {arXiv preprint arXiv:1708.05906}\
  } (\bibinfo {year} {2017})}\BibitemShut {NoStop}%
\bibitem [{\citenamefont {Waddington}\ \emph {et~al.}(2017)\citenamefont
  {Waddington}, \citenamefont {Sarracanie}, \citenamefont {Zhang},
  \citenamefont {Salameh}, \citenamefont {Glenn}, \citenamefont {Rej},
  \citenamefont {Gaebel}, \citenamefont {Boele}, \citenamefont {Walsworth},
  \citenamefont {Reilly} \emph {et~al.}}]{waddington17}%
  \BibitemOpen
  \bibfield  {author} {\bibinfo {author} {\bibfnamefont {D.~E.}\ \bibnamefont
  {Waddington}}, \bibinfo {author} {\bibfnamefont {M.}~\bibnamefont
  {Sarracanie}}, \bibinfo {author} {\bibfnamefont {H.}~\bibnamefont {Zhang}},
  \bibinfo {author} {\bibfnamefont {N.}~\bibnamefont {Salameh}}, \bibinfo
  {author} {\bibfnamefont {D.~R.}\ \bibnamefont {Glenn}}, \bibinfo {author}
  {\bibfnamefont {E.}~\bibnamefont {Rej}}, \bibinfo {author} {\bibfnamefont
  {T.}~\bibnamefont {Gaebel}}, \bibinfo {author} {\bibfnamefont
  {T.}~\bibnamefont {Boele}}, \bibinfo {author} {\bibfnamefont {R.~L.}\
  \bibnamefont {Walsworth}}, \bibinfo {author} {\bibfnamefont {D.~J.}\
  \bibnamefont {Reilly}},  \emph {et~al.},\ }\href@noop {} {\bibfield
  {journal} {\bibinfo  {journal} {Nature Communications}\ }\textbf {\bibinfo
  {volume} {8}} (\bibinfo {year} {2017})}\BibitemShut {NoStop}%
\bibitem [{\citenamefont {Yu}\ \emph {et~al.}(2005)\citenamefont {Yu},
  \citenamefont {Kang}, \citenamefont {Chang}, \citenamefont {Chen},\ and\
  \citenamefont {Yu}}]{yu05}%
  \BibitemOpen
  \bibfield  {author} {\bibinfo {author} {\bibfnamefont {S.-J.}\ \bibnamefont
  {Yu}}, \bibinfo {author} {\bibfnamefont {M.-W.}\ \bibnamefont {Kang}},
  \bibinfo {author} {\bibfnamefont {H.-C.}\ \bibnamefont {Chang}}, \bibinfo
  {author} {\bibfnamefont {K.-M.}\ \bibnamefont {Chen}}, \ and\ \bibinfo
  {author} {\bibfnamefont {Y.-C.}\ \bibnamefont {Yu}},\ }\href@noop {}
  {\bibfield  {journal} {\bibinfo  {journal} {Journal of the American Chemical
  Society}\ }\textbf {\bibinfo {volume} {127}},\ \bibinfo {pages} {17604}
  (\bibinfo {year} {2005})}\BibitemShut {NoStop}%
\bibitem [{\citenamefont {Chang}\ \emph {et~al.}(2008)\citenamefont {Chang},
  \citenamefont {Lee}, \citenamefont {Chen}, \citenamefont {Chang},
  \citenamefont {Tsai}, \citenamefont {Fu}, \citenamefont {Lim}, \citenamefont
  {Tzeng}, \citenamefont {Fang}, \citenamefont {Han} \emph {et~al.}}]{Chang08}%
  \BibitemOpen
  \bibfield  {author} {\bibinfo {author} {\bibfnamefont {Y.-R.}\ \bibnamefont
  {Chang}}, \bibinfo {author} {\bibfnamefont {H.-Y.}\ \bibnamefont {Lee}},
  \bibinfo {author} {\bibfnamefont {K.}~\bibnamefont {Chen}}, \bibinfo {author}
  {\bibfnamefont {C.-C.}\ \bibnamefont {Chang}}, \bibinfo {author}
  {\bibfnamefont {D.-S.}\ \bibnamefont {Tsai}}, \bibinfo {author}
  {\bibfnamefont {C.-C.}\ \bibnamefont {Fu}}, \bibinfo {author} {\bibfnamefont
  {T.-S.}\ \bibnamefont {Lim}}, \bibinfo {author} {\bibfnamefont {Y.-K.}\
  \bibnamefont {Tzeng}}, \bibinfo {author} {\bibfnamefont {C.-Y.}\ \bibnamefont
  {Fang}}, \bibinfo {author} {\bibfnamefont {C.-C.}\ \bibnamefont {Han}},
  \emph {et~al.},\ }\href@noop {} {\bibfield  {journal} {\bibinfo  {journal}
  {Nature nanotechnology}\ }\textbf {\bibinfo {volume} {3}},\ \bibinfo {pages}
  {284} (\bibinfo {year} {2008})}\BibitemShut {NoStop}%
\bibitem [{\citenamefont {Bumb}\ \emph {et~al.}(2013)\citenamefont {Bumb},
  \citenamefont {Sarkar}, \citenamefont {Billington}, \citenamefont
  {Brechbiel},\ and\ \citenamefont {Neuman}}]{Bumb13}%
  \BibitemOpen
  \bibfield  {author} {\bibinfo {author} {\bibfnamefont {A.}~\bibnamefont
  {Bumb}}, \bibinfo {author} {\bibfnamefont {S.~K.}\ \bibnamefont {Sarkar}},
  \bibinfo {author} {\bibfnamefont {N.}~\bibnamefont {Billington}}, \bibinfo
  {author} {\bibfnamefont {M.~W.}\ \bibnamefont {Brechbiel}}, \ and\ \bibinfo
  {author} {\bibfnamefont {K.~C.}\ \bibnamefont {Neuman}},\ }\href@noop {}
  {\bibfield  {journal} {\bibinfo  {journal} {Journal of the American Chemical
  Society}\ }\textbf {\bibinfo {volume} {135}},\ \bibinfo {pages} {7815}
  (\bibinfo {year} {2013})}\BibitemShut {NoStop}%
\bibitem [{\citenamefont {Liu}\ \emph {et~al.}(2007)\citenamefont {Liu},
  \citenamefont {Cheng}, \citenamefont {Chang},\ and\ \citenamefont
  {Chao}}]{liu07}%
  \BibitemOpen
  \bibfield  {author} {\bibinfo {author} {\bibfnamefont {K.-K.}\ \bibnamefont
  {Liu}}, \bibinfo {author} {\bibfnamefont {C.-L.}\ \bibnamefont {Cheng}},
  \bibinfo {author} {\bibfnamefont {C.-C.}\ \bibnamefont {Chang}}, \ and\
  \bibinfo {author} {\bibfnamefont {J.-I.}\ \bibnamefont {Chao}},\ }\href@noop
  {} {\bibfield  {journal} {\bibinfo  {journal} {Nanotechnology}\ }\textbf
  {\bibinfo {volume} {18}},\ \bibinfo {pages} {325102} (\bibinfo {year}
  {2007})}\BibitemShut {NoStop}%
\bibitem [{\citenamefont {Fu}\ \emph {et~al.}(2007)\citenamefont {Fu},
  \citenamefont {Lee}, \citenamefont {Chen}, \citenamefont {Lim}, \citenamefont
  {Wu}, \citenamefont {Lin}, \citenamefont {Wei}, \citenamefont {Tsao},
  \citenamefont {Chang},\ and\ \citenamefont {Fann}}]{fu07}%
  \BibitemOpen
  \bibfield  {author} {\bibinfo {author} {\bibfnamefont {C.-C.}\ \bibnamefont
  {Fu}}, \bibinfo {author} {\bibfnamefont {H.-Y.}\ \bibnamefont {Lee}},
  \bibinfo {author} {\bibfnamefont {K.}~\bibnamefont {Chen}}, \bibinfo {author}
  {\bibfnamefont {T.-S.}\ \bibnamefont {Lim}}, \bibinfo {author} {\bibfnamefont
  {H.-Y.}\ \bibnamefont {Wu}}, \bibinfo {author} {\bibfnamefont {P.-K.}\
  \bibnamefont {Lin}}, \bibinfo {author} {\bibfnamefont {P.-K.}\ \bibnamefont
  {Wei}}, \bibinfo {author} {\bibfnamefont {P.-H.}\ \bibnamefont {Tsao}},
  \bibinfo {author} {\bibfnamefont {H.-C.}\ \bibnamefont {Chang}}, \ and\
  \bibinfo {author} {\bibfnamefont {W.}~\bibnamefont {Fann}},\ }\href@noop {}
  {\bibfield  {journal} {\bibinfo  {journal} {Proceedings of the National
  Academy of Sciences}\ }\textbf {\bibinfo {volume} {104}},\ \bibinfo {pages}
  {727} (\bibinfo {year} {2007})}\BibitemShut {NoStop}%
\bibitem [{\citenamefont {Hovav}\ \emph {et~al.}(2010)\citenamefont {Hovav},
  \citenamefont {Feintuch},\ and\ \citenamefont {Vega}}]{Hovav10}%
  \BibitemOpen
  \bibfield  {author} {\bibinfo {author} {\bibfnamefont {Y.}~\bibnamefont
  {Hovav}}, \bibinfo {author} {\bibfnamefont {A.}~\bibnamefont {Feintuch}}, \
  and\ \bibinfo {author} {\bibfnamefont {S.}~\bibnamefont {Vega}},\ }\href@noop
  {} {\bibfield  {journal} {\bibinfo  {journal} {Journal of Magnetic
  Resonance}\ }\textbf {\bibinfo {volume} {207}},\ \bibinfo {pages} {176}
  (\bibinfo {year} {2010})}\BibitemShut {NoStop}%
\bibitem [{\citenamefont {Rej}\ \emph {et~al.}(2015)\citenamefont {Rej},
  \citenamefont {Gaebel}, \citenamefont {Boele}, \citenamefont {Waddington},\
  and\ \citenamefont {Reilly}}]{Rej15}%
  \BibitemOpen
  \bibfield  {author} {\bibinfo {author} {\bibfnamefont {E.}~\bibnamefont
  {Rej}}, \bibinfo {author} {\bibfnamefont {T.}~\bibnamefont {Gaebel}},
  \bibinfo {author} {\bibfnamefont {T.}~\bibnamefont {Boele}}, \bibinfo
  {author} {\bibfnamefont {D.~E.}\ \bibnamefont {Waddington}}, \ and\ \bibinfo
  {author} {\bibfnamefont {D.~J.}\ \bibnamefont {Reilly}},\ }\href@noop {}
  {\bibfield  {journal} {\bibinfo  {journal} {Nature communications}\ }\textbf
  {\bibinfo {volume} {6}} (\bibinfo {year} {2015})}\BibitemShut {NoStop}%
\bibitem [{\citenamefont {Chattergoon}\ \emph {et~al.}(2013)\citenamefont
  {Chattergoon}, \citenamefont {Mart{\'\i}nez-Santiesteban}, \citenamefont
  {Handler}, \citenamefont {Ardenkj{\ae}r-Larsen},\ and\ \citenamefont
  {Scholl}}]{Chattergoon13}%
  \BibitemOpen
  \bibfield  {author} {\bibinfo {author} {\bibfnamefont {N.}~\bibnamefont
  {Chattergoon}}, \bibinfo {author} {\bibfnamefont {F.}~\bibnamefont
  {Mart{\'\i}nez-Santiesteban}}, \bibinfo {author} {\bibfnamefont
  {W.}~\bibnamefont {Handler}}, \bibinfo {author} {\bibfnamefont {J.~H.}\
  \bibnamefont {Ardenkj{\ae}r-Larsen}}, \ and\ \bibinfo {author} {\bibfnamefont
  {T.}~\bibnamefont {Scholl}},\ }\href@noop {} {\bibfield  {journal} {\bibinfo
  {journal} {Contrast media \& molecular imaging}\ }\textbf {\bibinfo {volume}
  {8}},\ \bibinfo {pages} {57} (\bibinfo {year} {2013})}\BibitemShut {NoStop}%
\bibitem [{\citenamefont {Van~Criekinge}\ \emph {et~al.}(2011)\citenamefont
  {Van~Criekinge}, \citenamefont {Keshari}, \citenamefont {Vigneron},\ and\
  \citenamefont {Kurhanewicz}}]{van11}%
  \BibitemOpen
  \bibfield  {author} {\bibinfo {author} {\bibfnamefont {M.}~\bibnamefont
  {Van~Criekinge}}, \bibinfo {author} {\bibfnamefont {K.}~\bibnamefont
  {Keshari}}, \bibinfo {author} {\bibfnamefont {D.}~\bibnamefont {Vigneron}}, \
  and\ \bibinfo {author} {\bibfnamefont {J.}~\bibnamefont {Kurhanewicz}},\ }in\
  \href@noop {} {\emph {\bibinfo {booktitle} {Proceedings of the International
  Society for Magnetic Resonance in Medicine}}},\ Vol.~\bibinfo {volume} {19}\
  (\bibinfo {year} {2011})\ p.\ \bibinfo {pages} {1517}\BibitemShut {NoStop}%
\bibitem [{\citenamefont {Kucsko}\ \emph {et~al.}(2016)\citenamefont {Kucsko},
  \citenamefont {Choi}, \citenamefont {Choi}, \citenamefont {Maurer},
  \citenamefont {Sumiya}, \citenamefont {Onoda}, \citenamefont {Isoya},
  \citenamefont {Jelezko}, \citenamefont {Demler}, \citenamefont {Yao} \emph
  {et~al.}}]{kucsko16}%
  \BibitemOpen
  \bibfield  {author} {\bibinfo {author} {\bibfnamefont {G.}~\bibnamefont
  {Kucsko}}, \bibinfo {author} {\bibfnamefont {S.}~\bibnamefont {Choi}},
  \bibinfo {author} {\bibfnamefont {J.}~\bibnamefont {Choi}}, \bibinfo {author}
  {\bibfnamefont {P.~C.}\ \bibnamefont {Maurer}}, \bibinfo {author}
  {\bibfnamefont {H.}~\bibnamefont {Sumiya}}, \bibinfo {author} {\bibfnamefont
  {S.}~\bibnamefont {Onoda}}, \bibinfo {author} {\bibfnamefont
  {J.}~\bibnamefont {Isoya}}, \bibinfo {author} {\bibfnamefont
  {F.}~\bibnamefont {Jelezko}}, \bibinfo {author} {\bibfnamefont
  {E.}~\bibnamefont {Demler}}, \bibinfo {author} {\bibfnamefont {N.~Y.}\
  \bibnamefont {Yao}},  \emph {et~al.},\ }\href@noop {} {\bibfield  {journal}
  {\bibinfo  {journal} {arXiv preprint arXiv:1609.08216}\ } (\bibinfo {year}
  {2016})}\BibitemShut {NoStop}%
\bibitem [{\citenamefont {Panich}\ \emph {et~al.}(2015)\citenamefont {Panich},
  \citenamefont {Sergeev}, \citenamefont {Shames}, \citenamefont {Osipov},
  \citenamefont {Boudou},\ and\ \citenamefont {Goren}}]{Panich15}%
  \BibitemOpen
  \bibfield  {author} {\bibinfo {author} {\bibfnamefont {A.}~\bibnamefont
  {Panich}}, \bibinfo {author} {\bibfnamefont {N.}~\bibnamefont {Sergeev}},
  \bibinfo {author} {\bibfnamefont {A.}~\bibnamefont {Shames}}, \bibinfo
  {author} {\bibfnamefont {V.~Y.}\ \bibnamefont {Osipov}}, \bibinfo {author}
  {\bibfnamefont {J.}~\bibnamefont {Boudou}}, \ and\ \bibinfo {author}
  {\bibfnamefont {S.}~\bibnamefont {Goren}},\ }\href@noop {} {\bibfield
  {journal} {\bibinfo  {journal} {Journal of Physics: Condensed Matter}\
  }\textbf {\bibinfo {volume} {27}},\ \bibinfo {pages} {072203} (\bibinfo
  {year} {2015})}\BibitemShut {NoStop}%
\bibitem [{\citenamefont {van~der Wel}\ \emph {et~al.}(2006)\citenamefont
  {van~der Wel}, \citenamefont {Hu}, \citenamefont {Lewandowski},\ and\
  \citenamefont {Griffin}}]{Van06}%
  \BibitemOpen
  \bibfield  {author} {\bibinfo {author} {\bibfnamefont {P.~C.}\ \bibnamefont
  {van~der Wel}}, \bibinfo {author} {\bibfnamefont {K.-N.}\ \bibnamefont {Hu}},
  \bibinfo {author} {\bibfnamefont {J.}~\bibnamefont {Lewandowski}}, \ and\
  \bibinfo {author} {\bibfnamefont {R.~G.}\ \bibnamefont {Griffin}},\
  }\href@noop {} {\bibfield  {journal} {\bibinfo  {journal} {Journal of the
  American Chemical Society}\ }\textbf {\bibinfo {volume} {128}},\ \bibinfo
  {pages} {10840} (\bibinfo {year} {2006})}\BibitemShut {NoStop}%
\bibitem [{\citenamefont {Ajoy}\ \emph {et~al.}(2015)\citenamefont {Ajoy},
  \citenamefont {Bissbort}, \citenamefont {Lukin}, \citenamefont {Walsworth},\
  and\ \citenamefont {Cappellaro}}]{Ajoy15}%
  \BibitemOpen
  \bibfield  {author} {\bibinfo {author} {\bibfnamefont {A.}~\bibnamefont
  {Ajoy}}, \bibinfo {author} {\bibfnamefont {U.}~\bibnamefont {Bissbort}},
  \bibinfo {author} {\bibfnamefont {M.}~\bibnamefont {Lukin}}, \bibinfo
  {author} {\bibfnamefont {R.}~\bibnamefont {Walsworth}}, \ and\ \bibinfo
  {author} {\bibfnamefont {P.}~\bibnamefont {Cappellaro}},\ }\href {\doibase
  10.1103/PhysRevX.5.011001} {\bibfield  {journal} {\bibinfo  {journal} {Phys.
  Rev. X}\ }\textbf {\bibinfo {volume} {5}},\ \bibinfo {pages} {011001}
  (\bibinfo {year} {2015})}\BibitemShut {NoStop}%
\bibitem [{\citenamefont {Spence}\ \emph {et~al.}(2001)\citenamefont {Spence},
  \citenamefont {Rubin}, \citenamefont {Dimitrov}, \citenamefont {Ruiz},
  \citenamefont {Wemmer}, \citenamefont {Pines}, \citenamefont {Yao},
  \citenamefont {Tian},\ and\ \citenamefont {Schultz}}]{Spence01}%
  \BibitemOpen
  \bibfield  {author} {\bibinfo {author} {\bibfnamefont {M.~M.}\ \bibnamefont
  {Spence}}, \bibinfo {author} {\bibfnamefont {S.~M.}\ \bibnamefont {Rubin}},
  \bibinfo {author} {\bibfnamefont {I.~E.}\ \bibnamefont {Dimitrov}}, \bibinfo
  {author} {\bibfnamefont {E.~J.}\ \bibnamefont {Ruiz}}, \bibinfo {author}
  {\bibfnamefont {D.~E.}\ \bibnamefont {Wemmer}}, \bibinfo {author}
  {\bibfnamefont {A.}~\bibnamefont {Pines}}, \bibinfo {author} {\bibfnamefont
  {S.~Q.}\ \bibnamefont {Yao}}, \bibinfo {author} {\bibfnamefont
  {F.}~\bibnamefont {Tian}}, \ and\ \bibinfo {author} {\bibfnamefont {P.~G.}\
  \bibnamefont {Schultz}},\ }\href@noop {} {\bibfield  {journal} {\bibinfo
  {journal} {Proceedings of the National Academy of Sciences}\ }\textbf
  {\bibinfo {volume} {98}},\ \bibinfo {pages} {10654} (\bibinfo {year}
  {2001})}\BibitemShut {NoStop}%
\bibitem [{\citenamefont {Bouchard}\ \emph {et~al.}(2008)\citenamefont
  {Bouchard}, \citenamefont {Burt}, \citenamefont {Anwar}, \citenamefont
  {Kovtunov}, \citenamefont {Koptyug},\ and\ \citenamefont
  {Pines}}]{Bouchard08}%
  \BibitemOpen
  \bibfield  {author} {\bibinfo {author} {\bibfnamefont {L.-S.}\ \bibnamefont
  {Bouchard}}, \bibinfo {author} {\bibfnamefont {S.~R.}\ \bibnamefont {Burt}},
  \bibinfo {author} {\bibfnamefont {M.~S.}\ \bibnamefont {Anwar}}, \bibinfo
  {author} {\bibfnamefont {K.~V.}\ \bibnamefont {Kovtunov}}, \bibinfo {author}
  {\bibfnamefont {I.~V.}\ \bibnamefont {Koptyug}}, \ and\ \bibinfo {author}
  {\bibfnamefont {A.}~\bibnamefont {Pines}},\ }\href@noop {} {\bibfield
  {journal} {\bibinfo  {journal} {Science}\ }\textbf {\bibinfo {volume}
  {319}},\ \bibinfo {pages} {442} (\bibinfo {year} {2008})}\BibitemShut
  {NoStop}%
\bibitem [{\citenamefont {Kehayias}\ \emph {et~al.}(2017)\citenamefont
  {Kehayias}, \citenamefont {Jarmola}, \citenamefont {Mosavian}, \citenamefont
  {Fescenko}, \citenamefont {Benito}, \citenamefont {Laraoui}, \citenamefont
  {Smits}, \citenamefont {Bougas}, \citenamefont {Budker}, \citenamefont
  {Neumann} \emph {et~al.}}]{Kehayias17}%
  \BibitemOpen
  \bibfield  {author} {\bibinfo {author} {\bibfnamefont {P.}~\bibnamefont
  {Kehayias}}, \bibinfo {author} {\bibfnamefont {A.}~\bibnamefont {Jarmola}},
  \bibinfo {author} {\bibfnamefont {N.}~\bibnamefont {Mosavian}}, \bibinfo
  {author} {\bibfnamefont {I.}~\bibnamefont {Fescenko}}, \bibinfo {author}
  {\bibfnamefont {F.}~\bibnamefont {Benito}}, \bibinfo {author} {\bibfnamefont
  {A.}~\bibnamefont {Laraoui}}, \bibinfo {author} {\bibfnamefont
  {J.}~\bibnamefont {Smits}}, \bibinfo {author} {\bibfnamefont
  {L.}~\bibnamefont {Bougas}}, \bibinfo {author} {\bibfnamefont
  {D.}~\bibnamefont {Budker}}, \bibinfo {author} {\bibfnamefont
  {A.}~\bibnamefont {Neumann}},  \emph {et~al.},\ }\href@noop {} {\bibfield
  {journal} {\bibinfo  {journal} {Nat Commun}\ }\textbf {\bibinfo {volume}
  {8}},\ \bibinfo {pages} {188} (\bibinfo {year} {2017})}\BibitemShut {NoStop}%
\bibitem [{\citenamefont {Kaatze}\ \emph {et~al.}(2002)\citenamefont {Kaatze},
  \citenamefont {Behrends},\ and\ \citenamefont {Pottel}}]{Kaatze02}%
  \BibitemOpen
  \bibfield  {author} {\bibinfo {author} {\bibfnamefont {U.}~\bibnamefont
  {Kaatze}}, \bibinfo {author} {\bibfnamefont {R.}~\bibnamefont {Behrends}}, \
  and\ \bibinfo {author} {\bibfnamefont {R.}~\bibnamefont {Pottel}},\
  }\href@noop {} {\bibfield  {journal} {\bibinfo  {journal} {Journal of
  Non-Crystalline Solids}\ }\textbf {\bibinfo {volume} {305}},\ \bibinfo
  {pages} {19} (\bibinfo {year} {2002})}\BibitemShut {NoStop}%
\bibitem [{\citenamefont {Aslam}\ \emph {et~al.}(2013)\citenamefont {Aslam},
  \citenamefont {Waldherr}, \citenamefont {Neumann}, \citenamefont {Jelezko},\
  and\ \citenamefont {Wrachtrup}}]{Aslam13}%
  \BibitemOpen
  \bibfield  {author} {\bibinfo {author} {\bibfnamefont {N.}~\bibnamefont
  {Aslam}}, \bibinfo {author} {\bibfnamefont {G.}~\bibnamefont {Waldherr}},
  \bibinfo {author} {\bibfnamefont {P.}~\bibnamefont {Neumann}}, \bibinfo
  {author} {\bibfnamefont {F.}~\bibnamefont {Jelezko}}, \ and\ \bibinfo
  {author} {\bibfnamefont {J.}~\bibnamefont {Wrachtrup}},\ }\href@noop {}
  {\bibfield  {journal} {\bibinfo  {journal} {New Journal of Physics}\ }\textbf
  {\bibinfo {volume} {15}},\ \bibinfo {pages} {013064} (\bibinfo {year}
  {2013})}\BibitemShut {NoStop}%
\bibitem [{\citenamefont {Shim}\ \emph {et~al.}(2013)\citenamefont {Shim},
  \citenamefont {Nowak}, \citenamefont {Niemeyer}, \citenamefont {Zhang},
  \citenamefont {Brand{\~{a}}o},\ and\ \citenamefont {Suter}}]{Shim2013}%
  \BibitemOpen
  \bibfield  {author} {\bibinfo {author} {\bibfnamefont {J.~H.}\ \bibnamefont
  {Shim}}, \bibinfo {author} {\bibfnamefont {B.}~\bibnamefont {Nowak}},
  \bibinfo {author} {\bibfnamefont {I.}~\bibnamefont {Niemeyer}}, \bibinfo
  {author} {\bibfnamefont {J.}~\bibnamefont {Zhang}}, \bibinfo {author}
  {\bibfnamefont {F.~D.}\ \bibnamefont {Brand{\~{a}}o}}, \ and\ \bibinfo
  {author} {\bibfnamefont {D.}~\bibnamefont {Suter}},\ }\href
  {http://arxiv.org/abs/1307.0257} {\bibfield  {journal} {\bibinfo  {journal}
  {arXiv:1307.0257v2}\ } (\bibinfo {year} {2013})}\BibitemShut {NoStop}%
\bibitem [{\citenamefont {Ajoy}\ \emph {et~al.}(2017)\citenamefont {Ajoy},
  \citenamefont {Lv}, \citenamefont {Druga}, \citenamefont {Safvati},
  \citenamefont {Morabe}, \citenamefont {Fenton}, \citenamefont {Nazaryan},
  \citenamefont {Patel}, \citenamefont {Sjolander}, \citenamefont {Pagliero},
  \citenamefont {Reimer}, \citenamefont {Sakellariou},\ and\ \citenamefont
  {Meriles}}]{Ajoy17I}%
  \BibitemOpen
  \bibfield  {author} {\bibinfo {author} {\bibfnamefont {A.}~\bibnamefont
  {Ajoy}}, \bibinfo {author} {\bibfnamefont {X.}~\bibnamefont {Lv}}, \bibinfo
  {author} {\bibfnamefont {K.}~\bibnamefont {Druga}, \bibfnamefont
  {E.and~Liu}}, \bibinfo {author} {\bibfnamefont {B.}~\bibnamefont {Safvati}},
  \bibinfo {author} {\bibfnamefont {A.}~\bibnamefont {Morabe}}, \bibinfo
  {author} {\bibfnamefont {M.}~\bibnamefont {Fenton}}, \bibinfo {author}
  {\bibfnamefont {R.}~\bibnamefont {Nazaryan}}, \bibinfo {author}
  {\bibfnamefont {S.}~\bibnamefont {Patel}}, \bibinfo {author} {\bibfnamefont
  {T.}~\bibnamefont {Sjolander}}, \bibinfo {author} {\bibfnamefont
  {D.}~\bibnamefont {Pagliero}}, \bibinfo {author} {\bibfnamefont {J.~A.}\
  \bibnamefont {Reimer}}, \bibinfo {author} {\bibfnamefont {D.}~\bibnamefont
  {Sakellariou}}, \ and\ \bibinfo {author} {\bibfnamefont {A.}~\bibnamefont
  {Meriles}, \bibfnamefont {C.and~Pines}},\ }\href@noop {} {\bibfield
  {journal} {\bibinfo  {journal} {to be published}\ } (\bibinfo {year}
  {2017})}\BibitemShut {NoStop}%
\end{thebibliography}%
\bibliographystyle{apsrev4-1}

\pagebreak
\clearpage
\onecolumngrid
\begin{center}
\textbf{\large{\textit{Supplementary Information:} \\
\bluetitle{Orientation independent room-temperature optical $\Cs$ hyperpolarization in powdered diamond}}}\\
\hfill \break
\smallskip
\begin{minipage}[t]{0.625\textwidth}
\begin{center}
\end{center}
\end{minipage}
\end{center}


\twocolumngrid

\beginsupplement
{ \hypersetup{linkcolor=darkred}
\tableofcontents
}

\section{Methods and Materials}
\subsection{Surface area gains due to diamond powder}
\zsl{surface}

As presented in the main paper, the experiments in this work were performed with $200-250\mu$m diamond micro-particles from Element6 (Fig. 5C), with $\app$1ppm NV center concentration. The average edge length was found to be $87\pm3.9\ \mu$m and this measurement was used for all further calculations of surface area and volume.

Diamond powder has a surface area per unit volume orders of magnitude larger than a single crystal of equivalent mass, even considering special surface patterning~\cite{Kehayias17}.The surface area of the particles was calculated using $A=(6+12\sqrt{3})a^2$ where $a$ is the edge length. The average particle surface area was calculated to be $0.203\pm0.0004$ mm$^2$. Volume was calculated using the formula $V=8\sqrt{2}a^3$, yielding an average volume of $(7.45\pm0.00067)\times 10^{-3}$ mm\textsuperscript{3} and providing a surface area to volume ratio of 27.1 mm${}^{-1}$.

The mass of the entire sample of particles used in Fig. 2A of the main paper was $7.50\pm0.25$ mg. Using the known density of diamond (3.51 mg/mm\textsuperscript{3}), the total volume of sample was calculated to be $2.14\pm0.07$ mm\textsuperscript{3}. Dividing by the individual particle volume, the number of diamond particles was found to be $287\pm27$ diamonds, giving a total surface area of $58.3\pm5.1$ mm\textsuperscript{2}, equivalent to $7.77\pm0.72$ mm$^{2}$/mg. When compared to a diamond crystal of same mass and with dimensions 1mm$\times$1mm square base and 2.09 mm height, assuming only the 1 mm by 1 mm base makes contact with the liquid, the arrangement of many small particles results in a 58.3-fold increase in surface area of 7.77 mm$^{2}$/mg. In the case of diamond nanoparticles, with 100 nm edge lengths and a total volume of 2.14 mm${}^3$ and surface area of 6744 mm$^{2}$/mg, the increase in surface area would be more than 6700-fold. This dramatic increase in surface area would be highly beneficial in transferring spin polarization from diamond to a liquid sample. One method of lithographic application of nanogratings with 400 nm pitch with a depth of up to 3 micron increases the surface area of 2 mm x 2 mm x 0.5 mm diamond chips by 15, giving a surface area to mass ratio of $8.55$ mm$^2$/mg~\cite{Kehayias17}. Even with extensive processing, the resulting surface area is much less than that possible with 100 nm particles.

Our orientation independent technique can be used to hyperpolarize nano- and micro-particles of diamonds, both dry as well as suspended in solution (see Supplementary Material \zsr{solution}). We have been able to polarize particles of sizes $1$, $25$, $200$ and $400 \mu$m, and have found that the exact amount of polarization depends on optical penetration depth, concentration of P1 centers, and quality of sample preparation that sets the low field $T_1$.  We measured the $T_1$ times of the $200-250\mu$m particles to be 395.7s at $\T{B}_0=7$T and 10.19s at $\T{B}_{\R{pol}}=$8mT. Since the DNP polarization is inhomogeneous, i.e. spins closer to the NV centers are more strongly polarized, spin diffusion between them also causes a signal decay at low fields, leading to a super-exponential fall in signal. This feature slightly underestimates the $T_1$ at $\T{B}_{\R{pol}}$. A more detailed analysis of the spin diffusion factors affecting the $T_1$ will be presented elsewhere.

\begin{figure}[t]
  \centering
  {\includegraphics[width=0.49\textwidth]{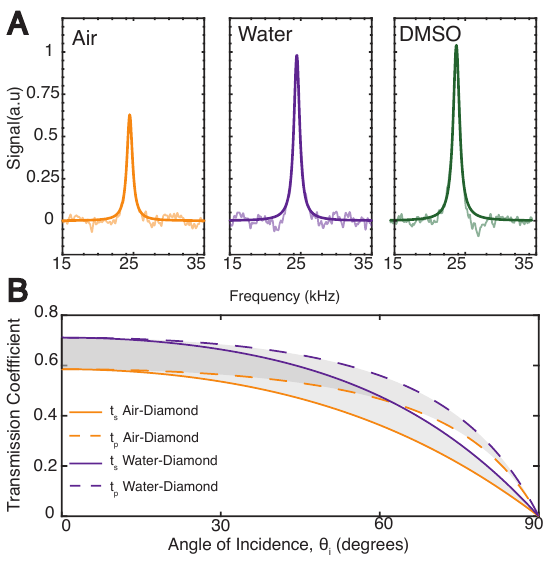}}
  \caption{\textbf{Enhanced $\Cs$ DNP for particles in solution.}  (A) Panels indicate that $\Cs$ polarization enhancements are in fact enhanced when the particles are immersed in solvents, which we attribute to better refractive index matching to diamond. We compare the obtained enhancements for e6 200$\mu$m diamond microparticles (Fig. 5C of main paper) dry ($n_r=1$), in water ($n_r=1.33$), and in DMSO ($n_r=1.48$). (B) The effect of better index matching obtained through Fresnel equations that relate the transmission coefficients between air-diamond and water-diamond interfaces. Here \I{s-} and \I{p-} refer to different polarizations, and for random diamond powder both contribute.}
 \zfl{index_matching}	
\end{figure}

\subsection{DNP of particles in solution}
\zsl{solution}
While in the paper we focus on \I{dry} particles in the sample tube, \zfr{index_matching} demonstrates the case when DNP is performed with the particles immersed in different solvents under the same conditions. The solvent, with no air bubbles, is held under slight pressure by the plunger (\zfr{DNP-setup}A inset). We find that the DNP enhancements are boosted by a factor $\app 2$, which we attribute potentially to better optical penetration into the diamonds due to better refractive index matching (\zfr{index_matching}B). We note that DNP performs without any visible sample heating inspite of microwave irradiation near $\app$ 2.45GHz, which is the operating frequency for conventional microwave ovens. This is likely due to water having low permittivity at that frequency~\cite{Kaatze02}.
 
\begin{figure}[t]
  \centering
  {\includegraphics[width=0.4\textwidth]{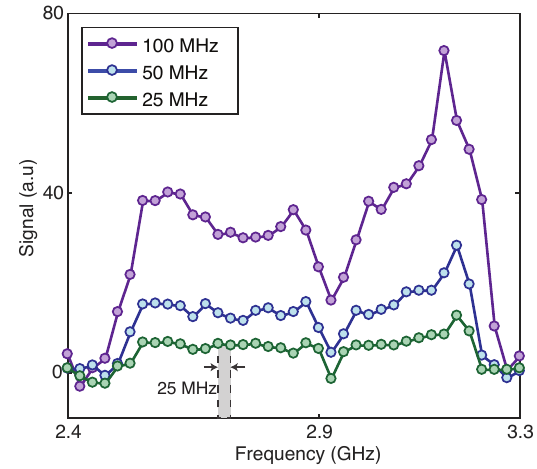}}
  \caption{\textbf{Electron powder pattern measured via $\Cs$ DNP.} Panel illustrates the obtained NV center powder pattern indirectly obtained by microwave sweeps of varying widths (in legend) according to Fig. 1B and measuring the $\Cs$ signal.  The obtained signal is a convolution of the underlying powder pattern with the sweep width. We find the same sign of the $\Cs$ hyperpolarization over the entire sweep, and remarkably that even rather large sweep widths eg. 100MHz can provide an faithful representation of the powder pattern.}
 \zfl{powder_bandwidths}	
\end{figure}

\begin{figure}[t]
  \centering
  {\includegraphics[width=0.4\textwidth]{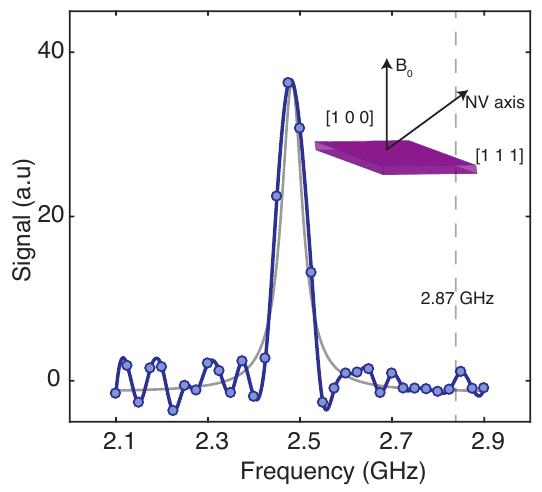}}
  \caption{\textbf{Single crystal electronic lineshape measured via $\Cs$ DNP.} Panel illustrates that by performing $\Cs$ DNP in a 25MHz window on a macroscopic single crystal (0.311 x 3.239 x 3.238 mm) sweeping from 2.4 to 2.7 GHz it is possible to map the underlying NV center electronic lineshape. In this case the magnetic field $\T{B}_{\R{pol}}\app$ 24.6mT is aligned at the magic angle 54.7$^{\circ}$ to the NV axis (inset) leading to the single spectral line for all four families of NV center orientations. Note that in these experiments the $\Cs$ spins indirectly report the electronic spectrum as opposed to direct ODMR measurements.}
 \zfl{single-crystal}	
\end{figure}

\section{Indirect NV Spectroscopy by $\Cs$ DNP}
\zsl{powder_pattern}
In the experiments in Fig. 3 of the main paper, we performed spectroscopy of the inhomogenously broadened NV center lineshape corresponding to the different NV orientations ( ``\I{powder pattern}'') indirectly. This was done by measuring the $\Cs$ enhancement sweeping microwaves through multiple 100MHz bandwidths, each bandwidth producing one data point. The relative signal obtained compared across multiple 100MHz windows reflects the powder pattern and is its convolution with the sweep bandwidth.

For clarity, and to quantify the effect of this convolution, in \zfr{powder_bandwidths} we repeat the experiment by varying the size of the sweep bandwidth at low field. The smallest window (25MHz) approaches the intrinsic linewidth of an individual NV center ($\app$ 10MHz). As expected the absolute value of the signal drops for smaller sweep windows, but the shape of the pattern remains qualitatively identical in each case. This proves convenient experimentally, because a quick map of the powder pattern allows one to optimize the full sweep bandwidth for maximum $\Cs$ polarization without ODMR or related optical spectroscopy.

\zfr{single-crystal} illustrates this also for a single crystal, where the $\Cs$ DNP is used to report on the underlying NV center spectral lineshape without ODMR. The experiment also demonstrates that our DNP mechanism is as effective on single crystals, and orientation independent (all four possible NV directions form a 54.7$^{\circ}$ angle with $B_0$ in \zfr{single-crystal}).

\section{Mechanism for orientation independent polarization transfer}
In this section, we outline the low-field DNP mechanism that governs the polarization transfer in our experiments. Before delving into the formal details, let us briefly summarize one more time its experimentally determined characteristics. 
\begin{enumerate}[label=(\roman*)]
\item The mechanism works predominantly at low field, 1-30mT. The exact field dependence has not been studied, however we find no DNP enhancement beyond 100mT.
\item The mechanism provides polarization transfer independent of the orientation of the NV center relative to the field direction, and subsequently allows the hyperpolarization of diamond powder (see \zfr{powder} of main paper, and \zfr{powder_bandwidths}).
\item The sign of the $\Cs$ hyperpolarization depends only on the direction of microwave sweep (see \zfr{results}C).
\item A low-to-high frequency sweep gives the nuclear spin polarized along the field (i.e. the same sign as thermal Boltzmann polarization).
\item Sweeping microwaves in the $m_s=-1$ or $m_s=+1$ NV manifold makes no difference to the obtained $\Cs$ hyperpolarization sign (\zfr{powder}A), which instead only depends on the MW sweep direction.
\item Even a $\lesssim$25MHz sweep window is enough to hyperpolarize spins (see \zfr{powder_bandwidths}). We expect the ultimate sweep window where hyperpolarization is achievable to be even smaller, but this was not experimentally verified due to lower absolute signal with decreasing window size. This points to the fact that the DNP mechanism does not rely only on closest shell $\Cs$ nuclei.
\item We experimentally find an optimal MW sweep rate of 40MHz/ms at a 24mT polarizing field (\zfr{results}D).
\item This optimal MW sweep rate is independent of laser power (\zfr{sweep_rate_laser}). 
\item The obtained enhancements are found to decrease with high laser power beyond a certain threshold $\app $2.5mW/mm$^2$ for 532nm excitation. Note that this number is also a function of the optical penetration depth in the sample, and potentially complex dynamics between the NV charge states~\cite{Aslam13}.
\item There is an optimal MW power (Rabi frequency), and increasing power leads to a drop in polarization transfer efficiency (\zfr{results}D inset).
\end{enumerate}

To model the observed spin dynamics, we use a two-spin system formed by the NV electronic spin and a hyperfine-coupled carbon, which is valid in the limit of a dilute $\Cs$ concentration (1$\%$ in the present case, the natural $\Cs$ abundance). The governing lab frame Hamiltonian is given by 
\beq\label{eq:1}
\begin{aligned}
	\mH = {} & \xD S_{z}^2 - \xg_e \vec{B} \cdot \vec{S} - \xg_C \vec{B} \cdot \vec{I} + A_{zz}S_{z}I_{z}\\
	& + A_{yy}S_{y}I_{y} + A_{xx}S_{x}I_{x} + A_{xz}S_{x}I_{z} + A_{zx}S_{z}I_{x}
\end{aligned}
\eeq
where $\vec{S} \equiv (S_{x},S_{y},S_{z})$ and $\vec{I} \equiv (I_{x},I_{y},I_{z})$ respectively denote the NV and $\Cs$ vector spin operators, $\xg_e$ ($\xg_C$) is the NV ($\Cs$) gyromagnetic ratio, $A_{\alpha\beta}$ with $\alpha,\beta = x,y,z$ are the components of the hyperfine tensor, and $\vec{B} \equiv (\sin\theta \ \cos\phi,sin\theta \ \sin\phi, cos\theta)$ is the magnetic field (10-30 mT) oriented along an axis characterized by a polar (azimuthal) angle $\theta$ ($\phi$) in a reference frame whose z-axis coincides with the NV direction; without loss of generality, we assume the $\Cs$ nucleus is contained within the xz-plane. Within the $m_S=\pm1$ states, the hyperfine coupling produces a $\Cs$ splitting
\beq\label{eq:2}
	\xo_{C}^{(\pm1)} = \sqrt{(A_{zz}\mp\xg_{C}B\cos\xt)^2 + A_{zx}^2}
\eeq
For the $m_S=0$ manifold, second-order perturbation theory leads to the approximate formula~\cite{Alvarez15,Shim2013},
\beq\label{eq:3}
\begin{aligned}
	\xo_{C}^{(0)} \approx {} & \xg_{C}B \\
	& + 2\Big(\frac{\xg_{e}B}{\xD}\Big) \sin\xt \big(\sqrt{A_{xx}^2 + A_{zx}^2} \cos^2\phi  + A_{yy}\sin^2\phi \big)
\end{aligned}
\eeq 
From Eqs.\ref{eq:2} and \ref{eq:3} we conclude that each manifold (including the $m_S=0$ manifold) has its own, distinct quantization axis which, in general, is different from the direction of the applied magnetic field. In particular, the second term in Eq. \ref{eq:3} can be dominant for hyperfine couplings as low as 1 MHz (corresponding to nuclei beyond the first two shells around the NV) if $\theta$ is sufficiently large, implying that, in general, $\Cs$ spins coupled to NVs misaligned with the external magnetic field experience a large frequency mismatch with bulk carbons, even if optical excitation makes $m_S=0$ the preferred NV spin state.

Assuming fields in the range 10-30 mT, it follows that $\Cs$ spins moderately coupled to the NV (300 kHz $\lesssim |A_{zz}|\lesssim$ 1 MHz) are dominant in the hyperpolarization process, not only because they are more numerous than those subject to strong couplings (i.e.,$|A_{zz}|\gtrsim$ 10 MHz corresponding to the first and second shells), but also because they more easily spin diffuse into the bulk (ultimately yielding the observable $\Cs$ signal). For sweep rates near the optimum ($\sim$ 40 MHz/ms), the time necessary to traverse the set of transitions connecting $m_S=0$ with either the $m_S=-1$ or $m_S=+1$ manifolds is relatively short \big($\lesssim$ 30 $\mu$s for weakly coupled carbons\big) meaning that optical repolarization of the NV preferentially takes place during the longer intervals separating two consecutive sweeps, as modeled in \zfr{mechanism} of the main text. 

The key to generating nuclear spin polarization is the selective non-adiabaticity of the mw sweep: Efficient polarization transfer takes place when the narrower Landau-Zener crossings connect branches with different electron and nuclear spin quantum numbers (\zfr{mechanism} in the main text), precisely the case in the $m_S=0\leftrightarrow m_S=-1$ ($m_S=0\leftrightarrow m_S=+1$) subset of transitions when the hyperfine coupling is positive (negative). For moderately weak couplings ($|A_{zz}|\lesssim$1 MHz) the hyperfine interaction has a stronger `dipolar-like' character (as opposed to `contact-like' character) implying that positive and negative couplings balance out. Therefore, when probing ensembles, both sets of transitions behave in the same way, i.e., $\Cs$ spins polarize positive in one direction, negative in the other (\zfr{powder}D in the main text). 

\zfr{simulations}A addresses the efficiency of the sweep mechanism in generating polarization as a function of the sweep velocity for two different hyperfine interactions and magnetic field alignments. For illustration purposes, we focus on the subset of transitions within the $m_S=0\leftrightarrow m_S=-1$ manifold and consider two cases where (i) the hyperfine interaction is $A_{zz} = A_{xx} = A_{yy} =$ 1 MHz and $(\theta, \phi) = (0^\circ,0^\circ)$, and (ii) the hyperfine interaction is $A_{zz} = A_{xx} = A_{yy} =$ 500 kHz with $(\theta, \phi) = (45^\circ,0^\circ)$. In either case, we sample the system evolution over a 20 MHz window respectively centered at 2590 MHz and 2672 MHz, which we sweep either from lower to higher frequencies or, conversely, from higher to lower frequencies (upper and lower halves of the plot in \zfr{simulations}A).

In the above simulation we assume the initial NV spin state is $m_S=0$ and consider $\Cs$ spins completely unpolarized. At the end of the sweep, we calculate the expectation value $\langle I_{z'} \rangle$ of nuclear spins in the direction of the external magnetic field. The polarization is then shown as a function of the sweep velocity. For simplicity, we neglect throughout these calculations the effect of NV spin-lattice relaxation. NV spin repolarization events due to optical pumping can take place randomly during the sweep but the outcome remains virtually unchanged if most take place away from the Landau-Zener crossing. As a matter of fact, NV spin repolarization is indeed crucial for resetting the population back into the $m_S=0$ subspace after the LZ passage, thus allowing for more polarization to be created in a second sweep. This argument holds for any subsequent sweeps and ultimately leads to a ratchet-like buildup of polarization that diffuses into the bulk. 

\zfr{simulations}B shows the calculated $\Cs$ polarization as a function of the Rabi field amplitude for a single sweep assuming a hyperfine coupling $A_{zz} = A_{xx} = A_{yy} =$ 500 kHz, a field direction $(\theta, \phi) = (45^\circ,0^\circ)$, and a sweep velocity $\dot{\omega} =$ 25 MHz/ms. Consistent with the observations in \zfr{results}D, we obtain an initial growth followed by a decay, with an optimum around 190 KHz. This response arises from the trade-off between the need to effectively drive the spin system throughout the full set of transitions and the widening gap between the two pairs of energy levels.   

A complete simulation including all angular coordinates $(\theta, \phi)$ and hyperfine interactions — required to quantitatively compare the numerical results with our experiments — is computationally-demanding and entails extensive additional work we will describe elsewhere. 

\begin{figure}[t]
  \centering
  {\includegraphics[width=0.483\textwidth]{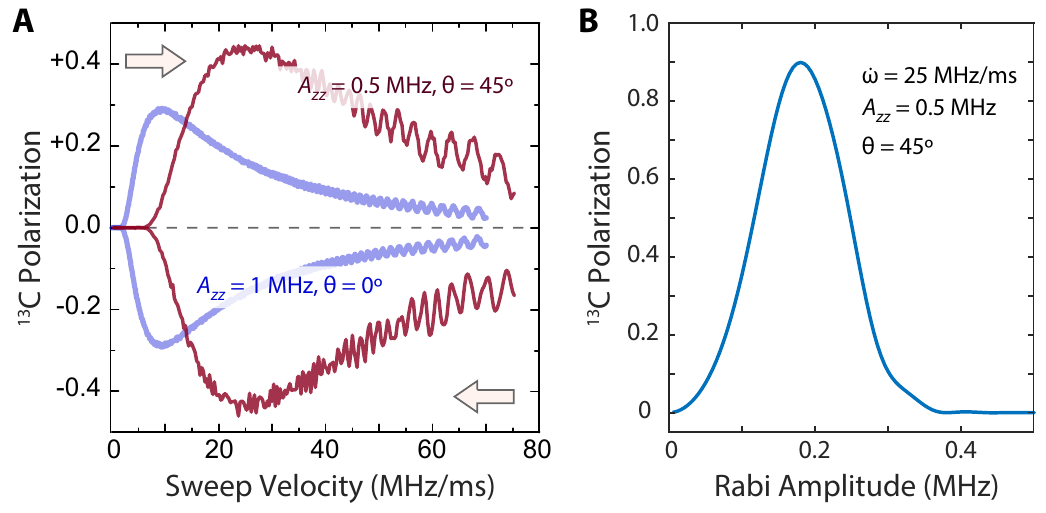}}
  \caption{
   \textbf{Simulations of $\Cs$ spin DNP enhancement.} (A) $\Cs$ spin polarization as a function of the sweep rate for two different magnetic field orientations and hyperfine couplings. Arrows indicate the direction of the sweep; low-to-high (high-to-low) frequency sweeps yield positive (negative) nuclear spin polarization (upper and lower halves of the plot, respectively). (B) Polarization efficiency as a function of the Rabi field amplitude for the listed hyperfine coupling and magnetic field direction. In (A) and (B) we use $A_{zx}=0.3A_{zz}$ and assume $A_{zz} = A_{xx} = A_{yy}$.
}
\zfl{simulations}
\end{figure}


\begin{figure*}{ht}
  \centering
	{\includegraphics[width=0.95\textwidth]{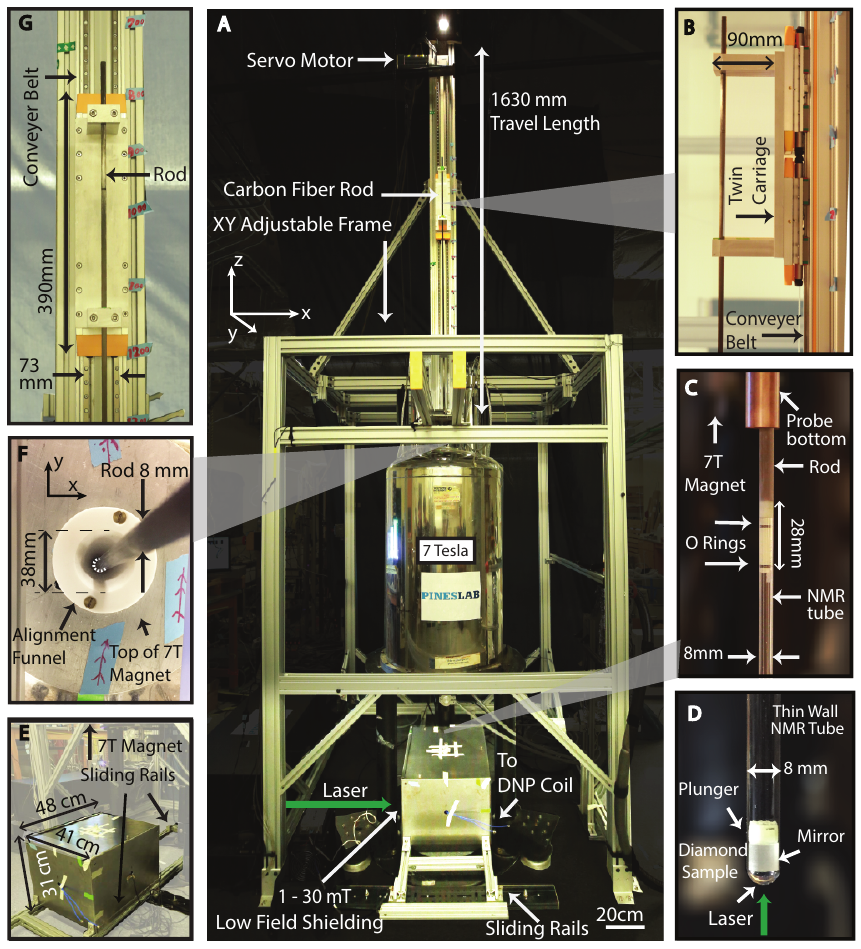}}
        \caption{\textbf{Detail of experimental setup} schematically described in Fig. 1A of main paper.   {(A)} Mechanical shuttler constructed over a 7T superconducting magnet. Polarization transfer of $\Cs$ in diamond particles is carried out by optical pumping at low field $\T{B}_{\R{pol}}$ (1-30mT) in a shielded volume below the magnet (\zfr{DNP-setup}), after which the sample is shuttled rapidly for measurement at 7T. A liquid nitrogen gun enables sample freezing at $\T{B}_{\R{pol}}$. {(B,G)} Shuttling is enabled by a carbon fiber rod that carries the sample. The rod is mounted on a movable twin carriage on the fast conveyer belt actuator stage.  {(C)} NMR tube that carries the sample is attached to the shuttling rod by a pressure fit arrangement using a pair of O-rings. {(D)} Diamond powder sample is contained with a dielectric mirror plunger employed to increase the efficiency of optical excitation. {(E)}  Iron shield producing $\T{B}_{\R{pol}}$ volume is placed on sliding rails to counter the magnetic force from the 7T magnet. {(F)} Bore of the 7T magnet is sealed with a teflon guide that ensures perfectly aligned shuttling and high fill-factor for inductive readout~\cite{Ajoy17I}.}
 \zfl{setup}	
\end{figure*}

\section{Experimental design}
\zsl{construction}
Our experimental setup consists of a fast field cycling device from 1mT-7T based on mechanical sample shuttling. The polarization transfer is performed at low field 1-30mT, after which the sample is rapidly shuttled to 7T for a high-sensitivity measurement of bulk $\Cs$ polarization. While this was schematically represented in Fig. 1A of the main paper, here we describe the actual device construction (\zfr{setup}). A more detailed analysis of the instrument will be presented elsewhere~\cite{Ajoy17I}.

\subsection{Construction and characteristics}
The field cycler consists of a XY-tunable shuttling tower constructed over a 7T superconducting magnet with a low field magnetic shield positioned below it (see \zfr{setup}). A conveyor belt driven actuator stage (Parker HMRB08) carries the sample along the fringing field of the magnet and into the shield. The sample is pressure-fit (\zfr{setup}C) onto a carbon-fiber shuttling rod (8mm, Rockwest composites) that is fastened rigidly on a twin-carriage mount on the actuator stage (\zfr{setup} A,B and G). The diamond powder is compactified by a plunger carrying a dielectric mirror (Thorlabs BB1-E02), which also serves to double pass the incoming laser radiation. The shuttler operates with a high positional precision of $50\mu$m, over a 1600mm travel length, at a maximum speed of 2m/s and acceleration of 30m/s$^{2}$. The shuttling from 7T-8mT takes $648\pm $2.5ms with a remarkably high repeatability over months of operation. The shuttling is triggered by a pulse generator (SpinCore PulseBlaster USB 100 MHz) via a high voltage MOSFET switch (Williamette MHVSW-001V-036V)~\cite{Ajoy17I}. The end of the motion triggers the NMR instrument for inductive detection.

To minimize vibration and obtain optimal sample fill factors for detection, we align the carbon fiber rod parallel to the magnet to better than 1 mdeg over the entire distance of travel. Dynamic alignment is performed by two funnel-shaped guiding stages made of soft teflon at the magnet bore (\zfr{setup}F), and the NMR probe. The NMR probe itself was designed hollow to accommodate fast sample shuttling to low fields below the magnet.  $\Cs$ detection at 75.03MHz was performed via printed saddle coils wrapped on a Quartz tube (9mm $\zt$ 11mm Technical Glass Products).

\begin{figure}[t]
  \centering
  {\includegraphics[width=0.49\textwidth]{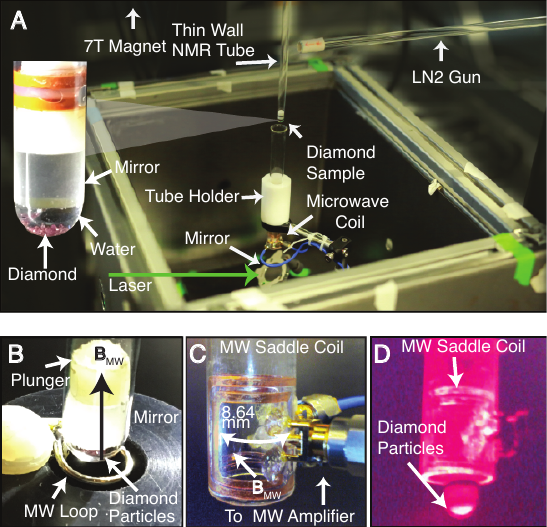}}
  \caption{\textbf{Low field DNP setup.} Figure illustrates the setup for polarization transfer at low field ($\T{B}_{\R{pol}}\sim$ 1-10 mT), performed inside a magnetic shield under the 7T NMR magnet. {(A)} The carbon fiber rod attached to the NMR tube carrying the diamond sample positioned inside the DNP excitation coil. Laser irradiation (5W over a 8 mm beam diameter) is applied from the bottom of the tube with a 45 degree mirror. \I{Inset:} Diamond particles immersed in water for better refractive index matching (\zfr{index_matching}). We use for MW excitation either (B) stub-loop antennas or (C) printed saddle coils providing MW fields $\T{B}_{\R{MW}}$ parallel or transverse to  $\T{B}_{\R{pol}}$. (D) Diamond particles under 1W laser illumination imaged with a 594nm long pass filter, showing the bright fluorescence from the NV centers. }
\zfl{DNP-setup}	
\end{figure}

\subsection{Low field DNP volume}
\zsl{low-field}
Polarization transfer from NV centers to $\Cs$ nuclei is carried out at low fields (1-30mT) in a magnetically shielded volume (\zfr{setup}E) below the 7T magnet.  The shield is constructed by concentric layers of stress annealed iron (NETIC S3-6 alloy 0.062'' thick, Magnetic Shield Corp.). Iron is used for its high magnetic saturation. The DNP field is tunable by varying position of the sample in the shield~\cite{Ajoy17I}. The shields are secured on sliding rails under the magnet (\zfr{setup}E) to contain upward forces. Laser light is irradiated to the bottom of the test tube holding the sample (\zfr{DNP-setup}). 200 mW of power provides 26.67 mW/mg to the diamond particles. Additionally, a MW antenna of 8.25 mm diameter applies swept-frequency microwave radiation that performs the polarization transfer (\zfr{DNP-setup}B,C).

\begin{figure*}[t]
  \centering
  {\includegraphics[width=0.96\textwidth]{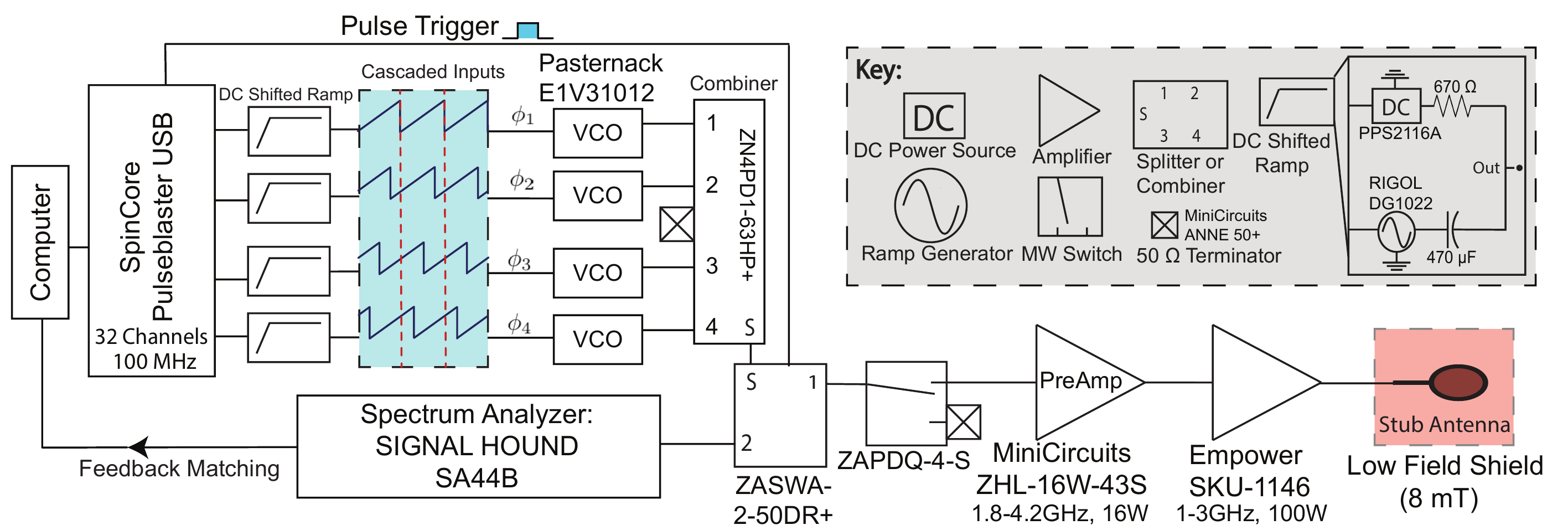}}
  \caption{\textbf{Schematic circuit for DNP excitation.} Low field DNP from NV centers to $\Cs$ is excited by microwave sweeps produced by employing voltage controlled oscillators (VCOs) with ramp generator inputs. Multiple VCOs are employed in a cascade to increase polarization transfer efficiency. A spectrum analyser is used to implement a feedback algorithm that exactly matches the VCO bandwidths to $<$2 MHz.  The microwaves are finally amplified by a 100W amplifier into a stub antenna that produces either longitudinal or transverse fields (\zfr{DNP-setup}) .}
\zfl{DNP-circuit}
\end{figure*}

\section{DNP Electronics setup}

Low field DNP is achieved by frequency sweeps over the NV center powder pattern during a continuous irradiation of laser light that polarizes the NV centers. Frequency sweeps are generated by a low-cost setup (\zfr{DNP-circuit}) involving voltage controlled oscillators (VCOs) -- we use either Pasternack PE1V31012 (1.6-3.2 GHz) or Minicircuits ZX95-3800A+ (1.9-3.7 GHz). An inexpensive arbitrary waveform generator (Rigol 1022A) is AC coupled to a programmable power supply (Circuit Specialists) to provide the ramp inputs to the VCOs to carry out the frequency sweeps. We combine the AC and DC sources in a high-pass configuration with 3dB cutoff frequency $\app 0.5$Hz, and both sources are individually USB controlled allowing us to calibrate the exact sweep bandwidth to a target range. The VCO output amplitude $\app $6dBm is controlled with a variable attenuator (Fairview SA1501SMA) to tune the MW Rabi frequency. 

The microwave outputs from the VCOs are subsequently amplified with a 100W amplifier (Empower SKU1146) and delivered to the DNP excitation coil. Different NV centers effectively see a different Rabi frequency based on their orientation with respect to the coil axis, and those with their N-to-V axis approximately parallel to this axis do not contribute to hyperpolarization. In our experiments, we employ with similar results both loop (\zfr{DNP-setup}B,) and saddle (\zfr{DNP-setup}C) type coil geometries with the microwave fields $\T{B}_{\R{MW}}$ parallel and perpendicular to the bias field $\T{B}_{\R{pol}}$. In principle, a pill-box geometry that combines both orientations will provide a gain in hyperpolarization enhancement by about a factor of 2 (\zfr{VCO-histogram}).

In order to increase the polarization transfer efficiency, we use a novel technique of cascaded microwave sweeps created via multiple VCOs. While we shall present a detailed analysis of the technique in a forthcoming publication, we present here the basic principle and implementation. The method works by effectively increasing the total number of microwave sweeps in a given total pumping time by cascading the ($N_{\R{VCO}}$). For an optimal cascaded sweep, the output microwave frequencies are time-shifted by $2\pi/N_{\R{VCO}}$ to maximize the period between successive sweeps. The DNP enhancement gains can be made to scale linearly with $N_{\R{VCO}}$ up to a certain limit set by the powder pattern (sweep bandwidth). With the simplicity of installing additional series of circuits, the addition of each VCO can be easily integrated into the design setup with the potential of additional sweepers allowing for large gains in polarization transfer efficiency.

The individual VCOs, even if of the same family, have slightly different frequency-voltage characteristics. To cascade the VCO sweeps effectively, we \I{match} their exact sweep bandwidths to within $\xD f<1$MHz via a fast-feedback mechanism employing  a spectrum analyzer (SignalHound SA44B) connected in parallel with the outputs (\zfr{DNP-circuit}). For matching, we employ a VCO sweep rate of 2kHz, allowing the output sweep spectrum to have a square-wave like shape, which is edge-detected in real time and feedback employed to converge to the target sweep bandwidth. The algorithm takes $\app$2 min to converge, and allows one to leverage several MW sweepers simultaneously, enabling a simple means to boost polarization transfer efficiency.

\begin{figure}[t]
  \centering
  {\includegraphics[width=0.44\textwidth]{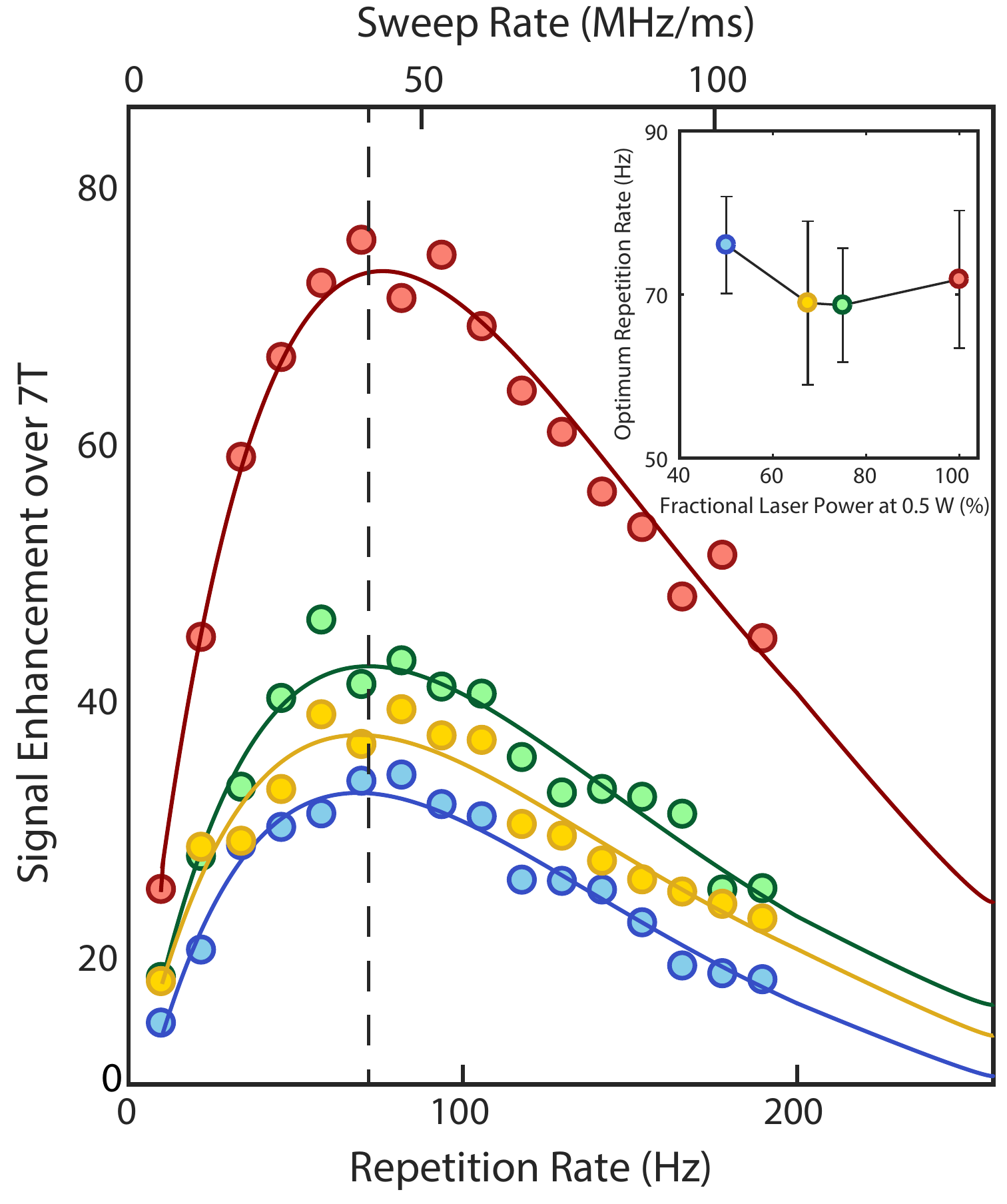}}
  \caption{\textbf{Sweep rate dependence on laser power.} (Main) $\Cs$ NMR signal as a function of the repetition (sweep) rate (lower and upper horizontal axes, respectively) for different illumination intensities. (Inset) The optimal repetition rate is nearly independent of laser intensity. In these experiments, the swept frequency bandwidth is 570 MHz and the laser beam diameter is 8 mm.}
 \zfl{sweep_rate_laser}	
\end{figure}

\section{DNP Optics setup}
\zsl{optics}
Collimated 532 nm laser light (Coherent Verdi) is employed for the optical polarization of NV center electrons, which is then subsequently transferred to the $\Cs$ nuclei by microwave irradiation. In practice, for the polarization of powders, the laser and swept microwaves are applied continuously (Fig. 1B of main paper). The laser power is controlled via a thin film polarizer and the half-wave plate (CVI TFP-532-PW-1025-C and QWPM-532-05-2, respectively) on a fine motorized rotation stage (Zaber RSW60A-T3). The laser beam is gated with an acousto-optic modulator (AOM, Isomet Corporation, M113-aQ80L-H), expanded to match the diameter of the NMR tube carrying the sample (8mm), and irradiated to the bottom of the tube. Fluorescence from the NV centers is imaged under a 594nm long pass filter (Semrock BLP01-594R-25) (\zfr{setup}D).

For most DNP experiments we operated with a total laser power of $\app$ 200mW over a 8mm beam diameter and 532nm excitation. We notice that the DNP efficiency decreases slightly at higher laser powers, which we attribute to the laser polarization breaking the coherence of the NV-$\Cs$ polarization transfer by repumping the NV center. \zfr{sweep_rate_laser} indicates that the optimal sweep rate employed for the microwave sweeps are independent of the laser power.

Let us now evaluate approximately the power employed per particle and per NV center. Assuming uniform distribution of laser power, each particle within the sample is hit with approximately 1mW. Using the total sample mass, the total number of particles, and the 1 ppm concentration of NV centers, it was determined that there are 1.31$\pm0.13\times$10$^{12}$ NV centers per diamond particle. As a result, the laser power to NV center ratio per particle is 7.64$\pm0.76\times$10$^{-13}$ mW/NV center. Overall this is a lower laser power than expected, and could indicate that the laser did not fully penetrate the diamond samples, thus reaching some NV centers and not others. Such a result would suggest potential for further enhancement if methods are developed to enhance laser penetration.

\begin{figure}[t]
  \centering
  {\includegraphics[width=0.49\textwidth]{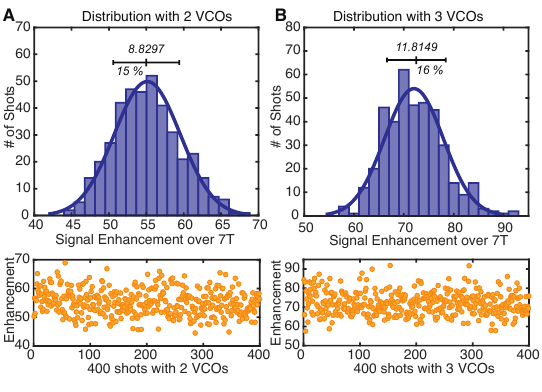}}
  \caption{\textbf{DNP enhancement spread due to orientational shaking.} Panels illustrate the spread in the obtained DNP enhancement factor of the $\Cs$ after 400 shots of the experiment measured at 7T for {(A)} 2 and {(B)} 3 VCOs sweeping the powder bandwidth.  The mean spread is approximately 15\% and we ascribe this to the orientational randomness of the powder for different experiment shots. It is also evident increasing the number of frequency sweepers leads to an enhanced DNP efficiency (\zfr{DNP-circuit}). }
\zfl{VCO-histogram}
\end{figure}

\section{Shot to shot variation of enhancement}

In our experiments, the DNP excitation coil producing the microwave sweeps is either transverse (saddle) or parallel (loop) to the 7T field. Moreover, when the particles are dry, they are free to re-orient, and do so due to the jerk experienced during fast shuttling.  This reorientation causes a shot-to-shot variation of the obtained enhancement (see \zfr{VCO-histogram}), since polarization cannot be transferred from NV centers that are parallel to the MW excitation. In future work, we plan to install a dual time-multiplexed MW excitation platform in both orthogonal directions parallel and transverse to the 7T field. This should allow a gain in average $\Cs$ polarization by approximately a factor of 2 (\zfr{VCO-histogram}). 

\section{Polarization loss due to shuttling}
In our experiments, since the polarization transfer is performed in a different (low) field $\T{B}_{\R{pol}}$ compared to where the $\Cs$ polarization is detected, the measurement in fact underestimates the polarization created due to the losses during sample shuttling. Note that for our goal of optically hyperpolarizing a liquid via the polarized $\Cs$ spins (Fig. 1C of main paper), we are concerned with the $\Cs$ polarization at $\T{B}_{\R{pol}}$, especially since the liquid $\Cs$ $T_1$ is high at low fields.

In this section, we quantify the shuttling loss by measuring the $T_1$ of the $\Cs$ spins as a function of position (field) and assuming a linear trajectory of motion. In total, the sample travels a distance of  $\Delta x = 829$mm to the NMR coil in $t_{\R{shuttle}} = 0.648$s. During this period, $\Cs$ spins relax at a rate $R_{z}(B) = 1/T_{1}(B)$, where $T_{1}(B)$ is the field dependent spin relaxation constant. For the e6 diamond microparticles (Fig. 5C) these range from 10.19s (8mT) to 395.7s (7T), and correspond to decays between $6.66\%$ and $.176\%$ respectively if static for $t_{\R{shuttle}}$ at these fields. The exact amount of polarization loss can be quantified as,

\beq
 \frac{dM_{z}(t)}{dt} = -R_{z}(M_{z}(t) - M_{z}^{0}) 
\zl{rate}
\eeq
where $M_{z}^{0}$ is the equilibrium magnetization. Integrating both sides, $\ln{\frac{(M_{z}(t) - M_{z}^{0})}{(M_{z}(0) - M_{z}^{0})}} = -\int_{0}^{t} \frac{1}{T_{1}(t)} dt$.

Using a numerical fit, we determine the equation of the experimentally obtained $T_1$ curve, $T_{1}(x) = 20.02423 e^{.003893x}$, where $x$ is the sample displacement from the low field position. Approximating the sample trajectory to be linear, we have $ x(t) \approx \frac{829}{.648}t = 1279.3t $. Now solving \zr{rate}, we estimate a polarization loss of only $\app 1.03\%$ during shuttling. For samples with shorter low-field relaxation times, this polarization loss is greater, and our measurement provides a larger underestimate of the polarization at $B_{\R{pol}}$.

\section{Data Processing}
In this section, we outline the procedure used for processing the $\Cs$ NMR data, and evaluating the enhancement factor obtained via our low-field DNP protocol. Enhancement factor is calculated by comparing the $\Cs$ signal at thermal equilibrium at 7T to our DNP signal. Because of the low inherent SNR, the thermal signal required 120 transients ($N_{\R{Thermal}}$) to obtain a clearly discernible signal.

For each data set, we appropriately phased and baseline-corrected each spectrum, and normalized the noise to unity. The enhancement was then calculated as the ratios of the normalized peak areas. To phase the spectra, we performed a zero-order phase correction that maximized the real portion of our peak by multiplying the real and imaginary portions of the peak by a phase value. The phasing algorithm detected both positive and negative peaks. Phased positive peaks had a reproducibly small range of phase values. Negative peaks, when phased to maximize the real portion of the peak, had phase values that were shifted by pi radians from the aforementioned range. By comparing measured phase values of a data set against the small range, the phasing algorithm detected whether the peak was positive or negative.

To correct the rolling baseline, we first used a peak detection algorithm. The peak detection algorithm fit an absorptive Lorentzian through the real portion of the peak, found the area of the entire fitted Lorentzian, and designated the signal limits such that the integral between the two limits was 90\% of the total area. After we found the peak limits, and consequently the noise sections, we fit a 12th order polynomial through both noise sections and subtracted the polynomial from the phased peak. This resulted in a peak with a flat baseline.

Subsequently, to quantify signal enhancement, we scaled each spectrum such that the noise normalized to unity. This allows signals taken with different number of averages are put on the same footing, allowing a convenient way to characterize enhancement gains due to hyperpolarization. The noise was defined to be the standard deviation of the non-peak sections of the spectra. After scaling both spectra, we found the area under each peak (corresponding to their respective SNRs) by calculating the Riemann sum from the left peak limit to the right peak limit, and calculated the signal enhancement with respect to 7T thermal equilibrium as,

\beq
\varepsilon = \fr{\R{SNR}_\R{DNP}}{\R{SNR}_\R{Thermal}}\:\sq{\fr{N_{\R{DNP}}}{N_{\R{Thermal}}}}
\eeq

At higher enhancements, the ratios of peak to noise are larger. Therefore phasing and baselining algorithms correct the peak heights proportionally to a greater extent compared to peaks of lower enhancement. This leads to the resulting curves being artificially non-smooth for certain data sets, which was not the case when the total absolute area was employed for each point. In order to account for this, we determined the ratio between the enhancement and the area for each data point, took the median of those ratios, and scaled the areas by the ratios to produce the final enhancement curves for all data sets.

\end{document}